\documentclass[reprint,nofootinbib]{revtex4}
\usepackage{graphicx}  
\usepackage{dcolumn}   
\usepackage{bm}        
\usepackage{amssymb}
\usepackage{hyperref}
\usepackage{multirow}
\usepackage{amsmath}
\usepackage{cases}
\usepackage{placeins}
\usepackage{amsfonts}

\usepackage{graphicx}
\usepackage{amsmath}
\usepackage{mathrsfs}
\usepackage[mathscr]{eucal}
\usepackage{array}
\usepackage[title]{appendix}
\usepackage{xcolor}
\usepackage{lipsum}
\usepackage{mathtools}
\usepackage{amssymb}
\usepackage{amssymb}
\usepackage{amsmath}
\usepackage{pifont}
\newcommand{\nn}{\nonumber}

\newcommand{\overbar}[1]{\mkern 1.2mu\overline{\mkern-1.5mu#1\mkern-1.5mu}\mkern 1.2mu}

\usepackage{epsfig}
\usepackage{makecell}
\usepackage{amsmath}
\usepackage{amsfonts}
\usepackage{amssymb}
\usepackage{mathtools}
\usepackage{url}
\usepackage{hyperref}
\usepackage{subfigure}
\usepackage{hhline}
\usepackage{color}
\usepackage{array,multirow}
\usepackage{bm}
\usepackage{cancel}
\newcommand{\beqa}{\begin{eqnarray}}
\newcommand{\eeqa}{\end{eqnarray}}
\newcommand{\beq}{\begin{equation}}
\newcommand{\eeq}{\end{equation}}

\newcommand{\bmt}{\begin{pmatrix}}
\newcommand{\emt}{\end{pmatrix}}
\usepackage{comment}
\newcommand{\be}{\begin{equation}}
\newcommand{\ee}{\end{equation}}
\newcommand{\bea}{\begin{eqnarray}}
\newcommand{\eea}{\end{eqnarray}}

\begin{document}
\title{Implications of light $Z'$ on semileptonic $B(B_{s}) \to T\{K_2^*(1430)$ $(f_2'(1525))\} \ell^+\ell^-$ decays at large recoil}
\author{Manas K. Mohapatra}
\email{manasmohapatra12@gmail.com}
\author{Anjan Giri}
\email{giria@iith.ac.in}                            
\affiliation{Department of Physics, IIT Hyderabad, Kandi - 502285, India}

\begin{abstract}
We probe the rare semileptonic decays $B_{(s)} \to K_2^*(1430)(f_2'(1525)) \ell^+\ell^-$ proceeding via $b\to s \ell \ell$ transition in the presence of a light $Z^{\prime}$ boson. We employ the presence of an additional vector type interaction and constrain the new physics coupling parameter using the existing experimental measurements on $R_K$ and $R_{K^*}$ observables. To understand the sensitivity of the new physics coupling, we investigate  the impact of this coupling on various physical observables such as differential branching ratio, the forward-backward asymmetry, the lepton polarization asymmetry, the angular observable $P_5^{\prime}$, and the lepton universality parameters such as the ratio of the branching ratio $R_{f_2'(K_2^*)}$ and some important Q parameters of $B_{(s)} \to K_2^*(1430)(f_2'(1525)) \ell^+\ell^-$ processes at large recoil. We find some  noticeable differences of the observables in the presence of light $Z^{\prime}$ contribution.
\end{abstract}
\maketitle

\section{Introduction}

According to our best understanding the standard model (SM), although a successful theory, is not enough to explain some key puzzles such as matter-antimatter asymmetry in the universe, dark matter, dark energy, hierarchy problem, neutrino mass and so on. Hunting for beyond the SM has been a challenge to the whole high energy physics community. To understand the nature, the flavor physics, in principle, could be the ideal platform to explore the new physics (NP) beyond the SM. In this respect the ongoing endevour in B meson decays are of great interest in testing the SM and shedding light the NP beyond it. However, in recent years, a few measurements in rare weak decays of B meson have shown deviations from the SM predictions both in flavor changing neutral current (FCNC) which undergo $b\to s \ell \ell$ parton level and in flavor changing charged current (FCCC) mediated by $b \to c \ell \nu$ transition. In the light of neutral quark level transitions, several measurements most importantly the lepton flavor universality violation (LFUV) parameter $R_{K^*}=\mathcal{BR} (B \to K^* \mu ^+ \mu ^-)/\mathcal{BR} (B \to K^* e^+ e^-)$ observed from LHCb~\cite{LHCb:2020lmf,LHCb:2017avl} and Belle~\cite{Belle:2019oag} have 2.1 - 2.4$\sigma$ deviation from SM prediction $\sim 1$~\cite{Bordone:2016gaq,Hiller:2003js}. However, recently the measurement of another clean observable $R_K=\mathcal{BR} (B \to K \mu ^+ \mu ^-)/\mathcal{BR} (B \to K e^+ e^-)$~\cite{Hiller:2014yaa, Hiller:2003js,Bordone:2016gaq} has been observed in the dilepton invariant mass-squared range range $ 1.1 \leq q^2 \leq 6.0$ $\rm GeV^2$ from LHCb experiment which indicates 3.1$\sigma$ discrepancy~\cite{LHCb:2021trn}. The experimental measurements of $R_K$ and $R_{K^*}$ are given as follows:

\bea
R_K^{Exp}&=& 0.846 \substack{+0.042 +0.013\\ -0.039-0.012}, \hspace{0.5cm} 1.1 \leq q^2\leq 6.0  \hspace{0.3cm}\rm GeV^2, \nn\\
R_{K^*}^{Exp}&=& 0.660 \substack{+0.11 \\ -0.07} \pm 0.03, \hspace{0.5cm} 0.045 \leq q^2\leq 1.1  \hspace{0.3cm} \rm GeV^2, \hspace{0.4cm} (\rm low \hspace{0.1cm} q^2) \nn\\
R_{K^*}^{Exp}&=& 0.690 \substack{+0.11 \\ -0.07} \pm 0.05, \hspace{0.5cm} 1.1 \leq q^2\leq 6.0 \hspace{0.3cm} \rm GeV^2, \hspace{0.4cm} (\rm central \hspace{0.1cm} q^2).
\eea

 Similarly another anomaly, so called the angular observable $P_5^{\prime}$ in $B\to K^* \mu^+ \mu^-$ decay mode observed from LHCb~\cite{LHCb:2013ghj,LHCb:2015svh}, ATLAS~\cite{ATLAS:2018gqc}, CMS~\cite{CMS Collaboration} and Belle~\cite{Belle:2016xuo} collaborations contribute (1 - 4)$\sigma$ deviations from the SM expectation~\cite{Descotes-Genon:2014uoa,Descotes-Genon:2013vna}. Further, a 3.6$\sigma$ deviation seen in the branching ratio of $B_s \to \phi \ell \ell$ process in the $q^2 \in$ [1.1,6.0] region by LHCb ~\cite{Bharucha:2015bzk,Aebischer:2018iyb}.
 
Decays of B meson to S-wave mesons (pseudo-scalar and vector mesons) have been explored widely both in theory as well as experiment, whereas the analysis of the P-wave mesons (scalar, axial vector and tensor mesons) in B decays have got relatively less attention. However, it is observed that a large amount of such decays have been established experimentally~\cite{ParticleDataGroup:2018ovx}. Therefore, in this work we intend to investigate the semileptonic decays of B meson into light P-wave tensor (T) mesons with $J^P=2^+$ containing $f_2^{\prime}(1525)$ and $K_2^*(1430)$ in the final state. The decay mode  
$B\to K_2^* \ell^+ \ell^-$  has been discussed in Ref.~\cite{Ahmed:2012zzc,RaiChoudhury:2006bnu,Hatanaka:2009gb,Hatanaka:2010fpr,Junaid:2011egj,Lu:2011jm,Aliev:2011gc,Das:2018orb}. Similarly, in Ref.~\cite{Li:2010ra}, though the authors  have investigated the NP effect in the presence of both vector like quark model and family non-universal $Z^{\prime}$ model, unfortunately a less emphasis was offered to $B_s \to f_2^{\prime} \ell^+ \ell^-$ process. However recently a detailed angular analysis of $B_s \to f_2^{\prime} \ell^+ \ell^-$ decay has been studied in the context of effective field theory framework~\cite{Rajeev:2020aut}. In this work we are not considering the branching ratios of $f_2^{\prime}$ and $K_2^*$ tensors in the given $B_s \to f_2^{\prime} \ell^+ \ell^-$ and $B\to K_2^* \ell^+ \ell^-$ processes, respectively. In the theoretical calculations, the knowledge of nonperturbative QCD is necessary which are parameterized in terms of decay constant, form factors. The form factors for $B_{(s)}\to T$ transition have been calculated in Isgur-Scora-Grinstein-Wise quark
model (ISGW)~\cite{Isgur:1988gb}  and in ISGW2 model~\cite{Scora:1995ty,Sharma:2010yx}, perturbative QCD method~\cite{Wang:2010ni} and light-cone sum rule (LCSR) approach~\cite{Yang:2010qd}.

Since the branching ratio includes the hadronic uncertainties unlike the clean observables $R_K$ and $R_{K^*}$, the NP is allowed in the muon and/or electron mode in $b\to s \ell^+ \ell^-$ quark level transition. Mostly in several works the authors have analyzed with heavy mediator such as heavy $Z^{\prime}$, leptoquarks~\cite{Alok:2017sui,Alok:2017jgr,Altmannshofer:2017yso,Hiller:2017bzc,Capdevila:2017bsm,Sala:2017ihs,Ciuchini:2017mik,DAmico:2017mtc,Hiller:2017bzc,Geng:2017svp,DiChiara:2017cjq} in the physics beyond the SM. However in the presence of light mediators, the discrepancy can also be explained for the observables like the $R_K$ and $R_{K^*}$~\cite{Datta:2017ezo,Datta:2017pfz,Ghosh:2017ber}. In this respect we consider a light $Z^{\prime}$ in which the NP Wilson coefficients are $q^2$ dependent~\cite{Sala:2017ihs,Ghosh:2017ber,Alok:2017sui,Bishara:2017pje,Datta:2017pfz} and study the impact on $B(B_{s}) \to T\{K_2^*(1430)$ $(f_2'(1525))\} \ell^+\ell^-$ decays.

The organization of the paper is as follows. In Section~\ref{kit}, we deliver the theoretical formalism that includes a brief review of generalized weak effective Hamiltonian for $b\to s \ell^+ \ell^-$ FCNC transition. Additionally we also present the $B \to T$ hadronic matrix elements. We provide the formulas of differential branching ratios and other observables of $B_s \to f_2^{\prime} \ell^+ \ell^-$ and $B\to K_2^* \ell^+ \ell^-$ processes in Sec.~\ref{obs}. In Sec.~\ref{NP}, we analyze the NP contribution in the presence of light $Z^{\prime}$ model. In Sec.~\ref{Results}, we discuss and analyze our results in the presence of new physics. To conclude, we provide a brief summary of our results in the Sec.~\ref{conc}.
\section{Formalism} \label{kit}
\subsection{Generalized effective weak Hamiltonian}
The generalized effective weak Hamiltonian for rare $b\to s \ell^+ \ell^-(|\Delta B|=|\Delta S| =1)$ transition is given as~\cite{Buras:1994dj,Misiak:1992bc}
\begin{eqnarray}
\label{eff_ham}
 \mathcal{H}_{eff} &=& - \frac{G_F}{\sqrt{2}}\, V_{tb}\, V_{ts}^{*}\, \frac{\alpha}{4\, \pi} \Bigg[ 
 {C}_{9}^{eff}\, \bar{s}\, \gamma^{\mu}\, P_L\, b\, \bar{l}\, \gamma_{\mu}\, l\, +\, 
 {C}_{10}^{eff}\, \bar{s}\, \gamma^{\mu}\, P_L\, b\, \bar{l}\, \gamma_{\mu}\, \gamma_{5}\, l\, -\, 
 \frac{2\,m_b}{q^2}{C}_{7}^{eff}\, \bar{s}\, i\, q_{\nu}\, \sigma^{\mu \nu}\, P_R\, b\, \bar{l}\, \gamma_{\mu}\, l\, \Bigg]\,,  
\end{eqnarray}
where $G_F$ is the Fermi coupling constant, $V_{ij}$ are the Cabbibo Kobayasi Maskawa (CKM) matrix element, $\alpha$ is the fine structure constant, $P_{L(R)}$ is the left (right) chiral project operator and $F_{\mu \nu}$ is the electromagnetic field strength tensor. The factorizable loop terms can be explained in terms of $C_7^{\rm eff}$ and $C_9^{\rm eff}$ as~\cite{Buras:1994dj}
\begin{eqnarray}
 {C}_{7}^{eff} &=& {C}_7 - \frac{{C}_5}{3} - {C}_6 \nonumber \\ 
 {C}_{9}^{eff} &=& {C}_{9}(\mu)\,+\,h(\hat{m}_c, \hat{s})\,{C}_{0}\, -\,\frac{1}{2}\,h(1,\hat{s})
 (4 {C}_{3}\,+\,4 {C}_{4}\,+\,3 {C}_{5}\,+\,{C}_{6})\, \nonumber \\ &&
 -\, \frac{1}{2}\,h(0,\hat{s}) ({C}_{3}\,+\, 3 {C}_{4})\, +\, \frac{2}{9}(3 {C}_{3}\,+ {C}_{4}\,+\,3 {C}_{5}\,+\,{C}_{6})\,,
\end{eqnarray}
where $\hat{m}_c=m_c/m_b$, $\hat{s}=q^2/m_{b}^2$,  and
${C}_0=3 {C}_1\, +\, {C}_2\, +\, 3{C}_3\, +\,{C}_4\, +\, 3{C}_5\, +\, {C}_6$.
The auxiliary functions given in the above equation are defined as
\begin{equation}
 h(z,\hat{s})=-\frac{8}{9}\ln \frac{m_b}{\mu}\,-\,\frac{8}{9}\ln z\, +\, \frac{8}{27}\, +\, \frac{4}{9}x\,-\,\frac{2}{9}(2+x)
 |1-x|^{1/2}\begin{cases}
             \ln \lvert \frac{\sqrt{1-x}+1}{\sqrt{1-x}-1}\rvert -i \pi\,, & \text{for $x \equiv \frac{4z^2}{\hat{s}}<1$} \\
             2 \arctan \frac{1}{\sqrt{x-1}}, & \text{for $x \equiv \frac{4z^2}{\hat{s}}>1$}
            \end{cases}
\end{equation}
\begin{equation}
 h(0,\hat{s})=-\frac{8}{9}\ln \frac{m_b}{\mu}\,-\,\frac{4}{9}\ln \hat{s}\, +\, \frac{8}{27}\, +\, \frac{4}{9} i \pi.
\end{equation}
The effective Wilson coefficient $C_9^{\rm eff}$ includes short distance contributions remain away from $c\bar{c}$ resonance zone whereas the long distance contributions which embed the resonant states [$J/{\psi}, \psi (2S),...$] from $b \to c\bar{c}s (\to s \ell^+ \ell^-)$ are excluded in our present analysis. Therefore we mainly dedicate to the $q^2$ rooms [0.045, 0.98] and [1.1, 6.0] $\rm GeV^2$ only. However we ignore the non factorizable corrections arising due to EMC (electromagnetic corrections) to the hadronic matrix elements in the effective Hamiltonian in this work. Moreover, the  $q^2$ dependent correction i.e the factorizable soft gluon part $\Delta C_9 (q^2)$ coming from charm loop effects are ignored in this work. However, the predicted ratio $\Delta C_9(q^2)/C_9$ has significant contribution to $B\to K \ell \ell$ and $B\to K^* \ell \ell$ which is $ \geq5 \%$ and reach upto $20 \%$, respectively\cite{Khodjamirian:2010vf}. In addition to this, recently in Ref.\cite{Gubernari:2020eft}, the authors have presented the non-local contributions to $b \to s$ transition modes, i.e, $B\to K^*$ and $B_s \to \phi$ decays where a modified analytic parameterization proposed in the non-local matrix elements.
However this is very difficult to calculate because it sign up the decay amplitude with non-perturbative non-local matrix elements. Therefore we don't consider this effect in this work.
\subsection{$B\to T(K_2^*(1430), f_2'(1525))$ hadronic matrix elements}
A tensor $T$ meson of spin-2 state polarization can be established in terms of spin-1 polarization vectors\cite{Berger:2000wt}. The given tensor can be written symbolically as $\epsilon ^{\mu \nu}(n)$ where $``n"$ correspond to $0, \pm 1,$ and $\pm 2$. The explicit expressions are given as follows\cite{Berger:2000wt,Li:2010ra,Wang:2010ni}:
\begin{eqnarray}
 \epsilon_{\mu\nu}(0)&=&\frac{1}{\sqrt{6}}\bigg[\epsilon_{\mu}(+)\, \epsilon_{\nu}(-)+\epsilon_{\nu}(+)\, \epsilon_{\mu}(-)\bigg]
 +\, \sqrt{\frac{2}{3}}\, \epsilon_{\mu}(0)\,\epsilon_{\nu}(0)\,,\nn\\
 \epsilon_{\mu\nu}(\pm1)&=&\frac{1}{\sqrt{2}}\bigg[\epsilon_{\mu}(\pm)\, \epsilon_{\nu}(0)+\epsilon_{\nu}(\pm)\, \epsilon_{\mu}(0)\bigg]
, \nonumber \\
 \epsilon_{\mu\nu}(\pm2)&=&\epsilon_{\mu}(\pm)\, \epsilon_{\nu}(\pm),
\end{eqnarray}
where
\begin{equation}
 \epsilon_{\mu}(0)=\frac{1}{m_T}(E_{T},0,0,\vec{p}_{T}), \hspace{1cm}
 \epsilon_{\mu}(\pm)=\frac{1}{\sqrt{2}}(0,\mp 1,-i,0)\,.
\end{equation}
Here $m_T$ is the mass, and $E_{T}$ and $\vec{p}_{T}$ are the energy and momentum of the tensor meson in the $B$  meson rest frame, respectively. However, the informations obtained from the helicity state for $n=2$ is not well understood of the final state two leptons. So the new polarization vector can be conveniently introduced as 
\begin{equation}
 \epsilon_{T_{\mu}}(h)=\frac{1}{m_{B_{(s)}}}\epsilon_{\mu\nu}(h)P_{B_{(s)}}^{\nu}\,,
\end{equation} 
where $P_{B_{(s)}}^{\nu}$ is the four momentum of the $B_{(s)}$ meson. The expressions of the new polarization vectors $\epsilon_{T_{\mu}}(h)$ ($h= 0, \pm1, \pm2$) are given explicitly as\cite{Li:2010ra}
\begin{eqnarray}
\epsilon_{T_{\mu}}(0) &=& \frac{1}{m_{B_{(s)}}}\sqrt{\frac{2}{3}}\,\epsilon(0)\cdotp P_{B_{(s)}}\epsilon_{\mu}(0)
 =\frac{\sqrt{\lambda}}{\sqrt{6}\,m_{B_s}m_T}\,\epsilon_{\mu}(0)\,\\ \nonumber
 \epsilon_{T_{\mu}}(\pm1) &=& \frac{1}{m_{B_{(s)}}}\frac{1}{\sqrt{2}}\,\epsilon(0)\cdotp P_{B_{(s)}}\epsilon_{\mu}(\pm)
 = \frac{\sqrt{\lambda}}{\sqrt{8}\,m_{B_{(s)}}m_T}\,\epsilon_{\mu}(\pm), \\ \nonumber
 \epsilon_{T_{\mu}}(\pm2) &=& 0,
\end{eqnarray}
where
\bea \label{lambda}
\lambda = m_{B_{(s)}}^4+ m_T^4+q^4-2(m_{B_{(s)}}^2 m_T^2+m_{B_{(s)}}^2q^2+q^2m_T^2).
\eea\\
The hadronic matrix elements of $B \to T$ transition, in analogy with $B \to V$, is given as\cite{Wang:2010ni,Yang:2010qd}
\begin{eqnarray}
 \langle\, T(P_{T}, \epsilon)|\bar{{(s)}}\gamma^{\mu}b|\bar{B}_{(s)} (P_{B_{(s)}})\,\rangle &=&
 -\frac{2V(q^2)}{m_{B_{(s)}}+m_{T}} \epsilon^{\mu \nu \rho \sigma}\, \epsilon_{T_{\nu}}^{*}\,P_{{B_s}\rho}P_{{T}\sigma} \nonumber \\ 
 \langle\, T(P_{T}, \epsilon)|\bar{s}\gamma^{\mu}\gamma_{5}b|\bar{B}_{(s)} (P_{B_{(s)}})\,\rangle &=&
 2 i\, m_{T}A_{0}(q^2)\frac{\epsilon_{T}^{*} \ldotp q}{q^2}\, q^{\mu}\, +\, i(m_{B_{(s)}}+m_{T})A_1 (q^2)
 \bigg[ \epsilon_{T_{\mu}}^{*}-\frac{\epsilon_{T}^{*} \ldotp q}{q^2}\, q^{\mu} \bigg]\, \nonumber \\ &&
 -\, iA_2 (q^2) \frac{\epsilon_{T}^{*} \ldotp q}{m_{B_{(s)}}+m_{f_{2}^{\prime}}}\, \bigg[ P^{\mu} - \frac{m_{B_{(s)}}^2 + m_{T}^2}{q^2}q^{\mu} \bigg] \nonumber \\
 \langle\, T(P_{T}, \epsilon)|\bar{s}\sigma^{\mu\nu}q_{\nu}b|\bar{B}_{(s)} (P_{B_{(s)}})\,\rangle &=&
 -2\,i\,T_1 (q^2)\, \epsilon^{\mu \nu \rho \sigma}\,\epsilon_{T_{\nu}}^{*}\,P_{{B_{(s)}}\rho}P_{{T}\sigma} \nonumber \\ 
 \langle\, T(P_{T}, \epsilon)|\bar{{(s)}}\sigma^{\mu\nu}\gamma_{5}q_{\nu}b|\bar{B}_{(s)} (P_{B_{(s)}})\,\rangle &=&
T_2(q^2)\, \bigg[(m_{B_{(s)}}^2 + m_{T}^2)\, \epsilon_{T_{\mu}}\,\epsilon_{T}^{*} \ldotp q\, P^{\mu}\bigg]\,
+\, T_3 (q^2)\, \epsilon_{T}^{*} \ldotp q\, \bigg[q^{\mu}-\frac{q^2}{m_{B_{(s)}}^2 + m_{T}^2}\,P^{\mu}\bigg]\,,
\end{eqnarray}
where the momentum transfer $q=P_{B_{(s)}} - P_T$. We use the relevant form factors in our analysis for $B_{(s)}$ to light $J^{PC}=2^{++}$ tensor meson (T) derived from the light-cone sum rule (LCSR) approach. 
The parameterized $q^2$ dependent form factors are given in the form as \cite{Yang:2010qd}:
\begin{equation}
 F^{B_{(s)}T}(q^2)=\frac{F^{B_{(s)}T}(0)}{1-a_T(q^2/m_{B_q}^2)+b_T(q^2/m_{B_q}^2)^2},
\end{equation}
where $F=V,A_0,A_1,A_2,T_1,T_2$ and $T_3$. The symbol $T$ denotes the tensor mesons $K_2^*(1430)$ and $f_2'(1525)$.
\section{Formulas of branching ratio and other observables} \label{obs}
The transition amplitude for $B \to K_2^*(1430) \ell^+\ell^-$ and $B_s \to f_2'(1525) \ell^+ \ell ^-$ processes can be obtained from the generalized effective Hamiltonian given in the Eq.~(\ref{eff_ham}). The $q^2$ dependent differential decay rate for the semileptonic $B_{(s)} \to T \ell^+ \ell^-$ $(T= f_2', K_2^*)$ modes mediated by $b\to s \ell ^+ \ell^-$ parton level can be given as~\cite{Zuo:2021kui,Li:2010ra,Rajeev:2020aut}
\bea
\frac{d\Gamma}{dq^2}=\frac{1}{4}(3I_1^c+6I_1^s-I_2^c-2I_2^s),
\eea
where the angular coefficients $I_i(q^2)$ are defined as
\bea
I_1^c&=&(|A_{L0}|^2+|A_{R0}|^2)+8\frac{m_\ell ^2}{q^2}Re[A_{L0}A_{R0}^*]+4\frac{m_\ell ^2}{q^2}|A_t|^2, \nn\\
I_1^s&=&\frac{3}{4}\big[|A_{L\perp}|^2+|A_{L\parallel}|^2+|A_{R\perp}|^2+|A_{R\parallel}|^2\big] \big(1-\frac{4m_\ell ^2}{3q^2}\big)+\frac{4m_\ell ^2}{q^2}Re\big[A_{L\perp}A_{R\perp}^*+A_{L\parallel} A_{R\parallel}^*\big],\nn\\
I_2^c&=&-\big(1-\frac{4m_\ell ^2}{q^2}\big)(|A_{L0}|^2+|A_{R0}|^2),\nn\\
I_{2}^{s} &=& \frac{1}{4} \big(1-\frac{4m_\ell ^2}{q^2}\big) \bigg[|A_{L\perp}|^2 + |A_{L\parallel}|^2 + |A_{R\perp}|^2 + |A_{R\parallel}|^2\bigg].
\eea
The explicit expressions of the transversity amplitudes given in the above equation can be written as follows:
\begin{eqnarray}
 A_{L0} &=& N_T\frac{\sqrt{\lambda(m_{(s)}^2,m_T^2,q^2)}}{\sqrt{6}\,m_{B_{(s)}}m_T}\frac{1}{2m_T\sqrt{q^2}}
 \bigg\{(C_{9}^{eff} - C_{10}) \bigg[(m_{B_{(s)}}^2 - m_T^2 - q^2)(m_{B_{(s)}} + m_T)A_1 - \frac{\lambda(m_{(s)}^2,m_T^2,q^2)}{m_{B_{(s)}} + m_T}A_2\bigg]+ \nonumber \\  
 && 2\,m_b\, C_{7}^{eff}\, \bigg[(m_{B_{(s)}}^2 + 3m_T^2 - q^2)T_2 - \frac{\lambda(m_{(s)}^2,m_T^2,q^2)}{m_{B_{(s)}}^2 - m_T^2}T_3 \bigg] \bigg\}\,, \nonumber \\
 A_{L\perp} &=& - N_T\sqrt{2} \frac{\sqrt{\lambda(m_{(s)}^2,m_T^2,q^2)}}{\sqrt{8}\,m_{B_{(s)}}m_T} \bigg[(C_{9}^{eff} - C_{10})
 \frac{\sqrt{\lambda(m_{(s)}^2,m_T^2,q^2)}}{m_{B_{(s)}} + m_T} V + \frac{\sqrt{\lambda(m_{(s)}^2,m_T^2,q^2)}\,2\,m_b\,C_{7}^{eff}}{q^2} T_1 \bigg]\,, \nonumber \\
 A_{L\parallel} &=& N_T\sqrt{2} \frac{\sqrt{\lambda(m_{(s)}^2,m_T^2,q^2)}}{\sqrt{8}\,m_{B_{(s)}}m_T} \bigg[(C_{9}^{eff} - C_{10})
 (m_{B_{(s)}} + m_T) A_1 + \frac{2\,m_b\,C_{7}^{eff}(m_{B_{(s)}}^2 - m_T^2)}{q^2} T_2 \bigg]\,, \nonumber \\
 A_{t} &=&2 N_T \frac{\sqrt{\lambda(m_{(s)}^2,m_T^2,q^2)}}{\sqrt{6}\,m_{B_{(s)}}m_T} C_{10} \frac{\sqrt{\lambda(m_{(s)}^2,m_T^2,q^2)}}{\sqrt{q^2}}A_0\,,\nn\\
 A_{Ri}&=&A_{Li}|_{C_{10}\to -C_{10}}, (i=0,\perp, \parallel),
\end{eqnarray}
where the normalization factor is given as
\begin{equation}
 N_T = \bigg[\frac{G_{F}^2\alpha^2}{3\cdot 2^{10}\pi^5\,m_{B_{(s)}}^3}|V_{tb}V_{ts}^{*}|^2 q^2 \sqrt{\lambda(m_{(s)}^2,m_T^2,q^2)}\bigg(1-\frac{4m_{l}^2}{q^2}
 \bigg)^{1/2} \bigg]^{1/2},
\end{equation}
and the parameter $\lambda$ is defined in the Eq.(\ref{lambda}).
Now in order to scrutinize the the structure of new physics, we explore with various interesting observables for the processes $B_{(s)}\to T \ell^+\ell^-$ and are given as follows~\cite{Zuo:2021kui}
\begin{itemize}
\item Differential branching ratio:
\bea
\mathcal{BR}(q^2)=\tau_{B_{(s)}}\frac{d\Gamma}{dq^2}=\tau_{B_{(s)}} \frac{1}{4}(3I_1^c+6I_1^s-I_2^c-2I_2^s)
\eea
\item Forward-backward asymmetry:
\bea
\langle \mathcal{A_{FB}}\rangle&=&\frac{\bigg(\int_0^1-\int_{-1}^0\bigg)d \cos \theta \frac{d^2\Gamma}{dq^2d \cos \theta}}{d\Gamma/dq^2}\nn\\
&=&\frac{3I_{6}}{3I_{1}^c + 6I_{1}^s - I_{2}^c -2I_{2}^s},
\eea
where \bea
I_6=2\sqrt{1-4m_\ell ^2/q^2} \big[Re(A_{L\parallel}A_{L\perp}^*)-Re(A_{R\parallel}A_{R\perp}^*)\big].
\eea
\item Longitudinal polarization fraction:
\bea
\langle \mathcal{F_L}\rangle=\frac{\int _{q_{low}^2}^{q_{high}^2}dq^2\frac{d\Gamma _L}{dq^2}}{\int _{q_{low}^2}^{q_{high}^2}dq^2\frac{d\Gamma}{dq^2}}=\frac{3I_{1}^c - I_{2}^c}{3I_{1}^c + 6I_{1}^s - I_{2}^c -2I_{2}^s}.
\eea
\item Angular observable $\langle \mathcal{P}_5^{\prime} \rangle$:
\bea
\langle \mathcal{P}_5^{\prime} \rangle= \frac{\int _{q_{low}^2}^{q_{high}^2} I_{5}}{2 \sqrt{-\int _{q_{low}^2}^{q_{high}^2}dq^2 I_{2}^c \int _{q_{low}^2}^{q_{high}^2}dq^2 I_{2}^s}}.
\eea
\end{itemize}
However, there are several other observables that can also be constructed and are very sensitive to the window of NP. These are defined in the form of ratios and differences between the observables associated with two different lepton families and are given explicitly as below.
\begin{itemize}
\item Lepton  flavor universality violation parameter:
\bea
\mathcal{R}_e^{\mu}(q_{low}^2,q_{high}^2)=\frac{\int _{q_{low}^2}^{q_{high}^2}dq^2 d\mathcal{BR}_{\mu}/dq^2}{\int _{q_{low}^2}^{q_{high}^2}dq^2 d\mathcal{BR}_e/dq^2}.
\eea
\item The $\langle Q_i \rangle$ $(i=\mathcal{F_L},\mathcal{A_{FB}}, Q_5^{\prime})$ parameter:
\bea
\langle Q_\mathcal{{F_L}}\rangle = \langle \mathcal{F}^\mu_\mathcal{L} \rangle - \langle \mathcal{F}^e_\mathcal{L} \rangle, \hspace{0.3cm} \langle Q_{\mathcal{A_{FB}}}\rangle = \langle \mathcal{A} ^ \mu_\mathcal{FB} \rangle - \langle \mathcal{A}^e_\mathcal{FB} \rangle, \hspace{0.3cm} \langle Q^{\prime}_5 \rangle = \langle Q^ \mu_5 \rangle - \langle Q^e_5 \rangle.
\eea
\end{itemize}
\section{New physics Analysis}\label{NP}
A heavy $Z^ \prime$ boson, in the tree level exchange with flavor changing neutral current transition mediated by $b \to s \ell^+ \ell ^- $ parton level, is the most obvious candidate in the NP contribution. There are different scenarios which are responsible for muonic four-fermion $b \to s \mu^+ \mu^-$ NP operators and are given as follows:
\bea
(\rm I):& &[\bar{s}\gamma _ \mu P_L b][\bar{\mu} \gamma ^ \mu \mu], \nn\\
(\rm II):& &[\bar{s}\gamma _ \mu P_L b][\bar{\mu} \gamma ^ \mu P_L \mu], \nn\\
(\rm III):& &[\bar{s}\gamma _ \mu \gamma _5 b][\bar{\mu} \gamma ^ \mu \mu].
\eea
However, the scenarios $\rm (I)$ and $\rm (II)$ display the $Z'$ boson to couple with the quark sector $\bar{s}_L-b_L-Z'$ and the lepton sector $Z'-\bar{\mu}-\mu$ vectorially whereas it couples axial-vectorially in the scenario  $\rm (III)$. Having said that we exclude the scenario $\rm (III)$  as it is strongly rejected by the $R_K$ measurement. The $Z'$ boson must transform as a singlet or triplet under $SU(2)_L$ gauge group as it couples to left handed quarks. In the case of triplet~\cite{Calibbi:2015kma,Crivellin:2015lwa,Boucenna:2016wpr}, a new gauge boson $W'$ can contribute to $B \to D^{(*)+}\tau ^- \bar{\nu}_ \tau$ mediated by $b \to c$ quark level transition where the deviation in the measurement has been observed in the Ref.~\cite{Belle:2015qfa,LHCb:2015gmp}. In the case of singlet under $SU(2)_L$ gauge group, this $Z'$ gauge boson associate with an extension of abelian $U(1)'$ group to the SM. 
Many works have been proposed in this model with the scenario $C_9 ^{\mu \mu}(NP)$ = -$C_{10} ^{\mu \mu}(NP)$ where the Wilson coefficients are $q^2$-independent. However, on the other hand, it is very interesting to consider a light $Z'$ which can also address $b\to s \mu^+ \mu^-$ data~\cite{Datta:2017pfz,Datta:2017ezo, Alok:2017sui}. If $2 m_ \mu <m_{Z'}< m_B$, a resonance state can be obtained in the dimuon invariant mass. Moreover to say that since no signature for such kind of state has been observed in the dimuon invariant mass, we consider the typical $Z'$ mass less than $2m_ \mu$ i.e 200 $\rm MeV$ in our analysis. For the coupling $\bar{s}b$ with the light $Z'$, the general form of the flavor changing vertex $\bar{s}bZ'$ is considered as~\cite{Alok:2017sui}
\bea
F(q^2) \bar{s}\gamma ^\mu P_L b Z_ \mu ^{\prime},
\eea
where the form of the form factor $F(q^2)$ can be written as
\bea
F(q^2)=a_L^{bs}+g_L^{bs}\frac{q^2}{m_B^2}+...
\eea
The leading order term $a_L^{bs}$ given in the above equation is severely constrained by $B\to K \nu \bar{\nu}$ and can be neglected, and we consider the coupling $g_L^{bs}$ only. Thus the $q^2$-dependent NP Wilson coefficients for $b\to s \mu^+ \mu^-$ transition are given as
\bea
C_9^{\mu \mu}(NP)= \mathcal{G} \frac{g^L_{bs}q^2/m_B^2(g^L_{\mu \mu}+g^R_{\mu \mu})}{q^2-m_{Z'}^2}, \hspace{0.5cm}C_{10}^{\mu \mu}(NP)= - \mathcal{G} \frac{g^L_{bs}q^2/m_B^2(g^L_{\mu \mu}-g^R_{\mu \mu})}{q^2-m_{Z'}^2},
\eea
where $\mathcal{G}=\frac{\pi}{\sqrt{2}G_F \alpha V_{tb}V_{ts}^*}$.
 It has been pointed out in Ref.~\cite{Alok:2017sui} that one can explain the B anomalies as good as in the case of heavy $Z'$ boson. It is clearly reported that except $R_{K^*}$ measurement in the low $q^2$ bin range, the light $Z'$ with pure vector coupling to muon can easily accommodate the clean observables $R_K^{[1,6]}$ and $R_{K^*}^{[1.1,6]}$ data given in Table - I of Ref.~\cite{Datta:2017ezo}. Since we assume the NP exist in muonic mode of $b\to s 
\ell^+ \ell^-$ transition, the NP coupling $C_9^{\mu \mu}(NP)$ is considered in our analysis where the light $Z'$ couple with muon vectorially under the condition $g^L_{\mu \mu}=g^R_{\mu \mu}=g_{\mu \mu}$. 

The long-standing discrepancy between theory and experiment that concerns with the anomalous magnetic dipole moment of muon i.e  $a_\mu = (g-2)/2$ has put an excitement among theorists. The combination of the recent updates on the measurements from Fermilab~\cite{Muong-2:2021ojo} and the previous result obtained from Brookhaven National Laboratory E82~\cite{Muong-2:2006rrc} leads a new average value with 4.2 $\sigma$ deviation from the SM result~\cite{Aoyama:2020ynm} and are given as follows:
\bea
a_\mu ^{SM}&=&116591810(43)\times 10^{-11}, \hspace{0.5 cm} a_\mu ^{exp}=116592061(43)\times 10^{-11},\nn\\
\Delta a_\mu &\equiv & a_\mu ^{exp}-a_\mu ^{SM} = (2.51 \pm 0.59) \times 10^{-9}.
\eea
As the light $Z'$ can also explain the muon $(g-2)$ anomaly, from the Ref.~\cite{Leveille:1977rc} the expression of the absolute magnitude of the discrepancy $\Delta a_ \mu$ is given as
\bea
\Delta a_ \mu =\frac{(g_{\mu \mu})^2}{8 \pi ^2} \int _0^1 \frac{2x^2(1-x)}{x^2+(m_{Z'}^2/m_ \mu ^2)(1-x)}dx,
\eea
where $m_{Z'}$ is the mass of light $Z'$ boson, $m_ \mu$ is the mass of muon and the coupling $g_{\mu \mu} =1.42 \times 10^{-3}$ is obtained for $m_{Z'}=200$ MeV. 
\section{Results and discussion}\label{Results}
\subsection{Relevant input parameters}
In this subsection, we report all the relevant inputs used for the numerical calculations of the various decay observables. In our analysis the input parameters such as mean life time and masses of $B_{(s)}$, the tensor mesons and lepton masses, the Fermi coupling constant are given as follows~\cite{ParticleDataGroup:2020ssz}
\bea
\tau_{B} &=&1.638 \times 10^{-12}\hspace{0.1cm}{\rm sec},\hspace{0.1cm} m_B=5.27934\hspace{0.1cm} {\rm GeV},\hspace{0.1cm} m_{B_s}=5.36688\hspace{0.1cm}{\rm GeV}, \nn\\
\tau_{B_s} &=&1.515 \times 10^{-12}\hspace{0.1cm}{\rm sec},\hspace{0.1cm} m_{K_2^*}=1.430\hspace{0.1cm}{\rm GeV}, \hspace{0.1cm}m_{f_2^{\prime}}= 1.525\hspace{0.1cm}{\rm GeV},\nn\\
G_F&=&1.1663787 \times 10^{-5}\hspace{.1cm}{\rm GeV^{-2}},\hspace{0.1cm} m_e=0.5109989461\times 10^{-3}\hspace{0.1cm}{\rm GeV},\hspace{0.1cm} m_\mu =0.1056583715 \hspace{0.1cm} {\rm GeV}.
\eea
Similarly for the quark masses, we use $m_b^{\rm pole}=4.8 \hspace{0.1cm}{\rm GeV}, m_b^{(\overbar{\rm MS})}=4.2\hspace{0.1cm}{\rm GeV}$, and $m_c^{(\overbar{\rm MS})}=1.28\hspace{0.1cm} {\rm GeV}$ \cite{Altmannshofer:2008dz}. From the Ref. \cite{ParticleDataGroup:2020ssz}, we also consider the fine structure constant $\alpha =1/133.28$ and the CKM parameter $|V_{tb}V_{ts}|=0.04088(55)$. The inputs of the Wilson coefficients in the leading logarithm approximation calculated at $\mu =4.8$ are taken from the Ref. \cite{Ali:1999mm} and are given in Table \ref{tab_wc}.

\begin{table}[htbp]
\centering
\setlength{\tabcolsep}{8pt} 
\renewcommand{\arraystretch}{1.5} 
\scalebox{1.3}{
\begin{tabular}{ccccccccc}
\hline
\hline
$C_1$ & $C_2$ & $C_3$ & $C_4$ & $C_5$ & $C_6$ & $C_{7}^{eff}$ & $C_9$ & $C_{10}$ \\
\hline
-0.248 & 1.107 & 0.011 & -0.026 & 0.007 & -0.031 & -0.313 & 4.344 & -4.669 \\
\hline 
\hline
\end{tabular}
}
\caption{Wilson coefficients $C_{i}(m_b)$ in the leading logarithmic approximation~\cite{Ali:1999mm}}
\label{tab_wc}
\end{table}

\begin{table}[htbp]
\centering
\setlength{\tabcolsep}{5pt} 
\renewcommand{\arraystretch}{1.3} 
\scalebox{1.2}{
\begin{tabular}{ccc}
\hline
\hline
 & $[F^{BK_2^*}(0),a_{K_2^*},b_{K_2^*}]$ &$[F^{B_sf_2^{\prime}}(0),a_{f_2^{\prime}},b_{f_2^{\prime}}]$\\
\hline
$V$ & $[0.16\pm0.02,2.08,1.50]$ &$[0.15\pm0.02,2.06,1.49]$\\ 
$A_0$ & $[0.25\pm0.04, 1.57,0.10]$ &$[0.25\pm0.04,1.72,0.31]$\\
$A_1$ & $[0.14\pm0.02,1.23,0.49]$ &$[0.13\pm0.02,1.25,0.47]$ \\
$A_2$ & $[0.05\pm0.02,1.32,14.9]$ &$[0.03\pm0.02,4.71,105]$\\
$T_1$ & $[0.14\pm0.02,2.07,1.50]$ &$[0.13\pm0.02,2.06,1.49]$\\
$T_2$ & $[0.14\pm0.02,1.22,0.35]$ &$[0.13\pm0.02,1.23,0.32]$\\
$T_3$ & $[0.01\substack{+0.02\\-0.01},9.91,276]$ &$[0.00\substack{+0.02\\-0.01},\text{--},\text{--}]$\\
\hline 
\hline
\end{tabular}
}
\caption{The relevant form factors with the fitted parameters~\cite{Yang:2010qd}}
\label{tab_ff}
\end{table}

However we report the relevant form factors required for the computation of the decay observables from the Ref. \cite{Yang:2010qd}. The explicit entries of the form factors at $q^2=0$ with the fitted parameters $a$ and $b$ are given in Table - \ref{tab_ff}.
\subsection{$\chi ^2$ analysis}
To obtain the discrepancy of the SM with the experimental data, we perform a naive $\chi ^2$ analysis with the existing $b\to s \ell \ell$ data. In our fit, we only include the updated experimental result obtained from LHCb for $R_K^{[1.1,6.0]}$~\cite{LHCb:2021trn} and $R_{K^*}^{[1.1,6.0]}$~\cite{LHCb:2017avl} in our analysis as the $R_{K^*}$ measurement in the bin range 0.045 $\leq q^2\leq$ 1.1 $\rm GeV^2$ does not accommodate within 1$\sigma$ deviation. The $\chi ^2$ is defined as
\bea
\chi^2= \sum_i  \frac{\Big ({\cal O}_i^{\rm th}(C_9^{\rm NP}) -{\cal O}_i^{\rm Exp} \Big )^2}{(\Delta {\cal O}_i^2)},
\eea
where the numerator includes the theoretical contributions $\mathcal{O}_i^{\rm th}$ with the NP coupling and the measured central values $\mathcal{O}_i^{\rm Exp.}$ of the observables, and $\Delta {\cal O}_i^2=(\Delta {\cal O}_i^{\rm Exp})^2+(\Delta {\cal O}_i^{\rm SM})^2$. The denominator envelop $1 \sigma$ uncertainties from theory and experimental results. Considering the coupling as real, we obtain the best fit value of the NP coupling associated with the $Z'$ boson as $g_{bs}^L=1.57 \times 10^{-5}$.

\subsection{$B \to K_2^*(1430) \ell^+\ell^-$  and $B_s \to f_2'(1525) \ell^+\ell^-$ decay observables}
\subsubsection*{Analysis of $B_s \to f_2^{\prime}(1525) \ell^+ \ell^-$ in SM and beyond}
\begin{figure}[htb]
\centering
\setlength{\tabcolsep}{8pt} 
\renewcommand{\arraystretch}{1.5}
\includegraphics[scale=0.7]{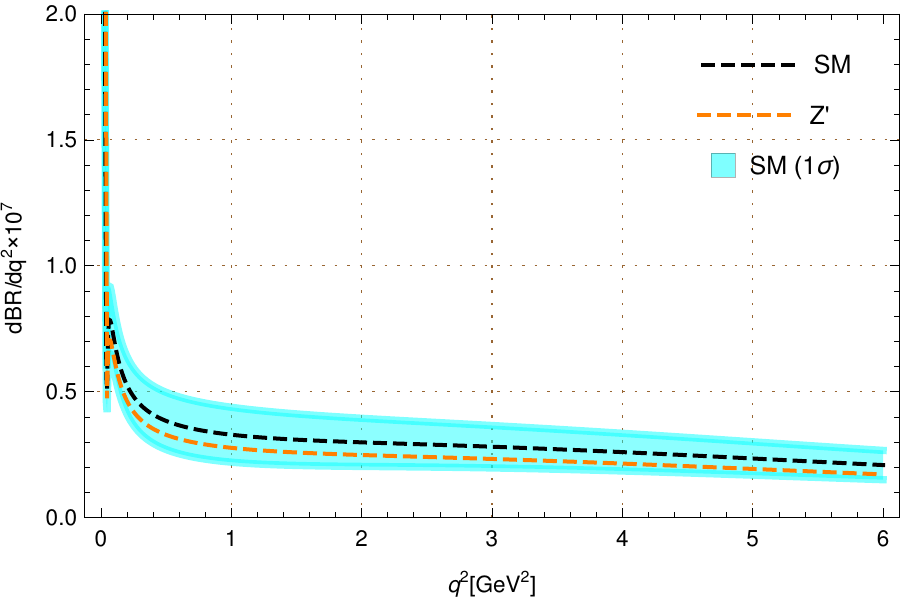} 
\quad
\includegraphics[scale=0.7]{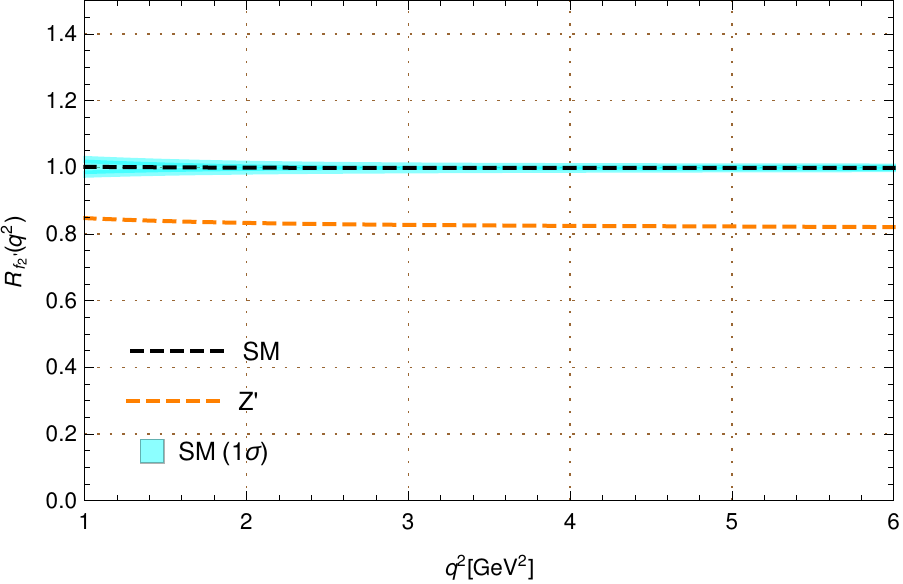} 
\quad
\includegraphics[scale=0.7]{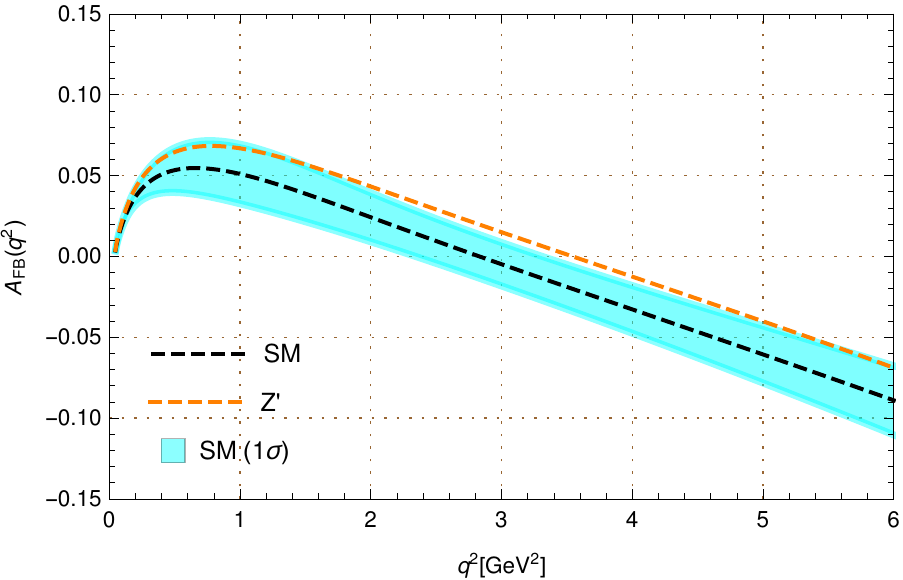} 
\quad
\includegraphics[scale=0.7]{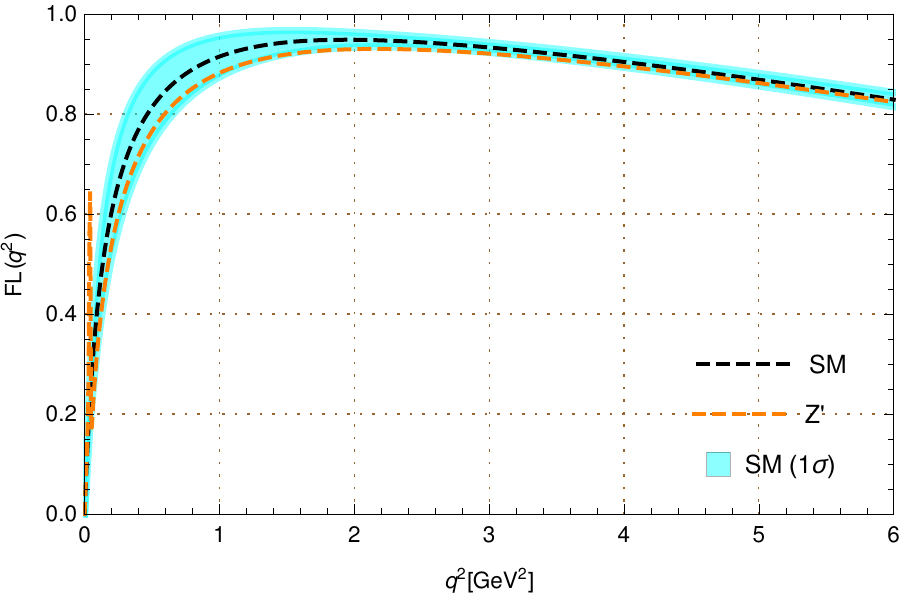}
\quad
\includegraphics[scale=0.7]{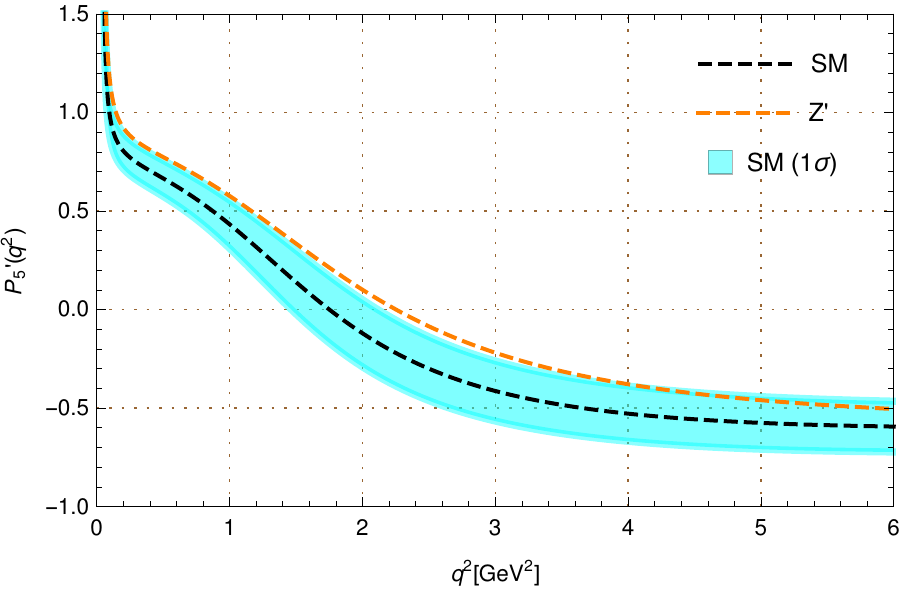}
\caption{\label{Bf2_Fig1}The $q^2$ distribution of various observables such as branching ratio, the polarization fraction, the forward-backward asymmetry and $P_5^{\prime}$ for $B_s \to f_2^{\prime}(1525) \ell ^+ \ell ^-$ process (black dotted line: SM contribution, cyan band: $1 \sigma$ uncertainty due to form factors and CKM element, orange dotted line: $Z^{\prime}$ contribution).}
\end{figure}

\begin{figure}[htbp]
\centering
\includegraphics[scale=0.63]{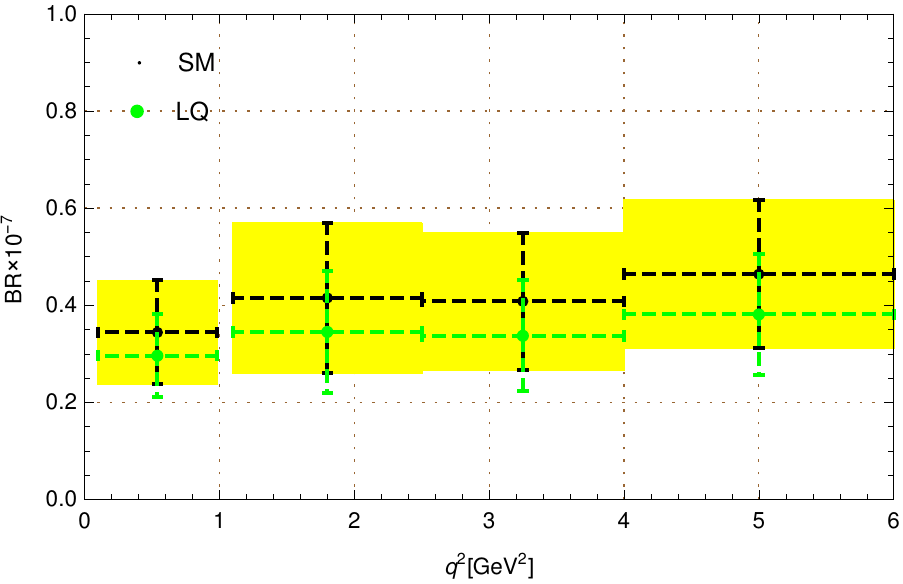} 
\includegraphics[scale=0.63]{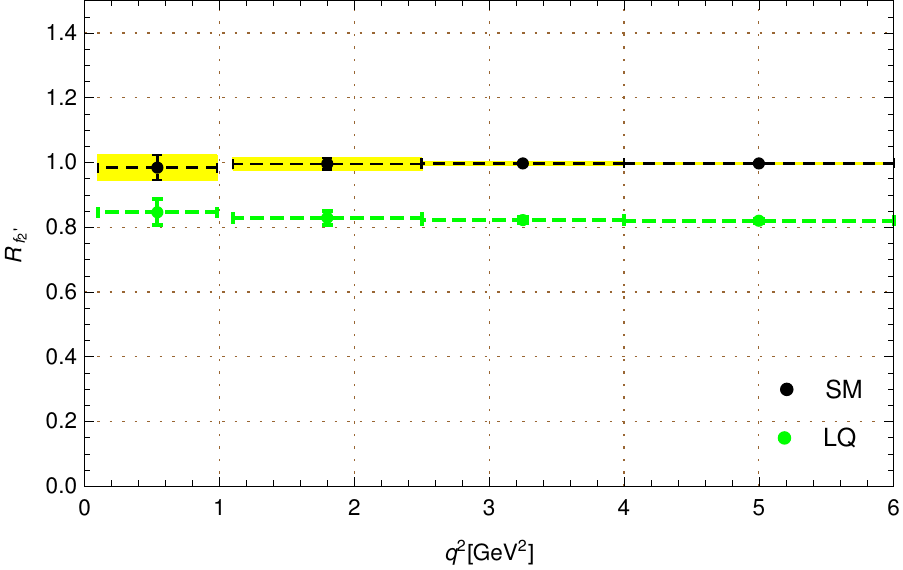} 
\includegraphics[scale=0.63]{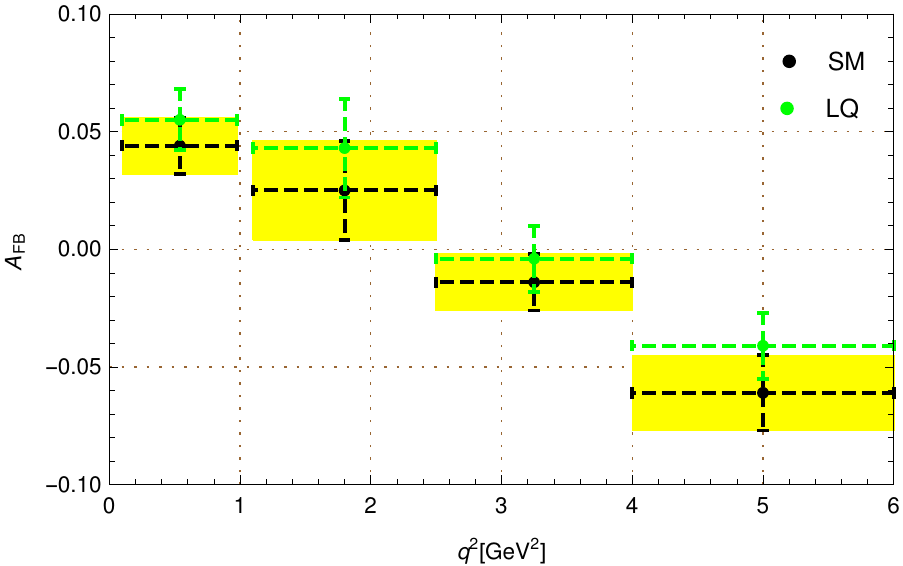} 
\includegraphics[scale=0.63]{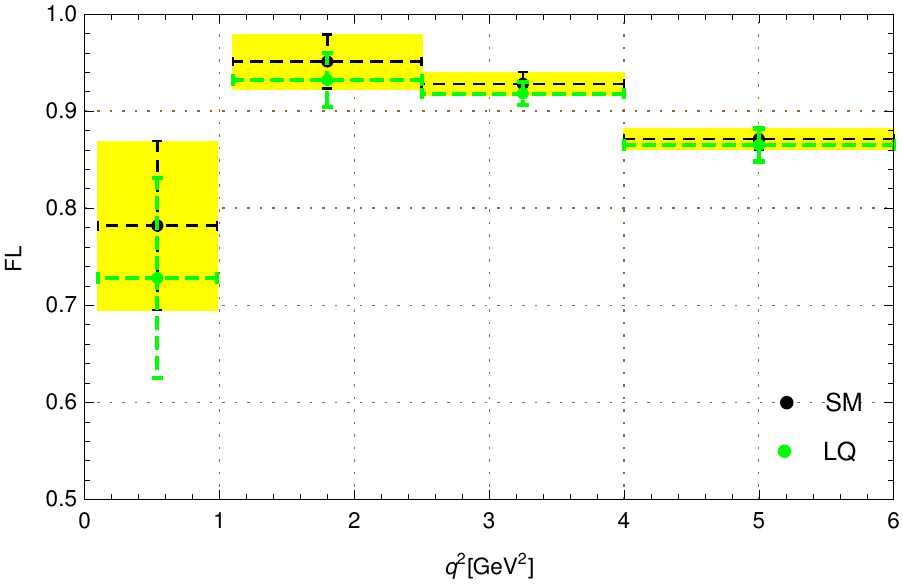}
\includegraphics[scale=0.63]{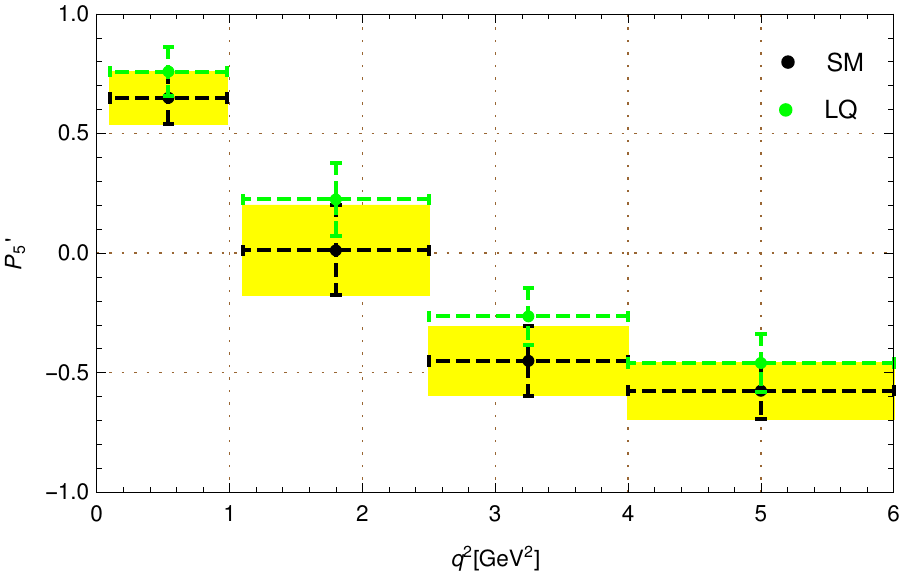}

\includegraphics[scale=0.63]{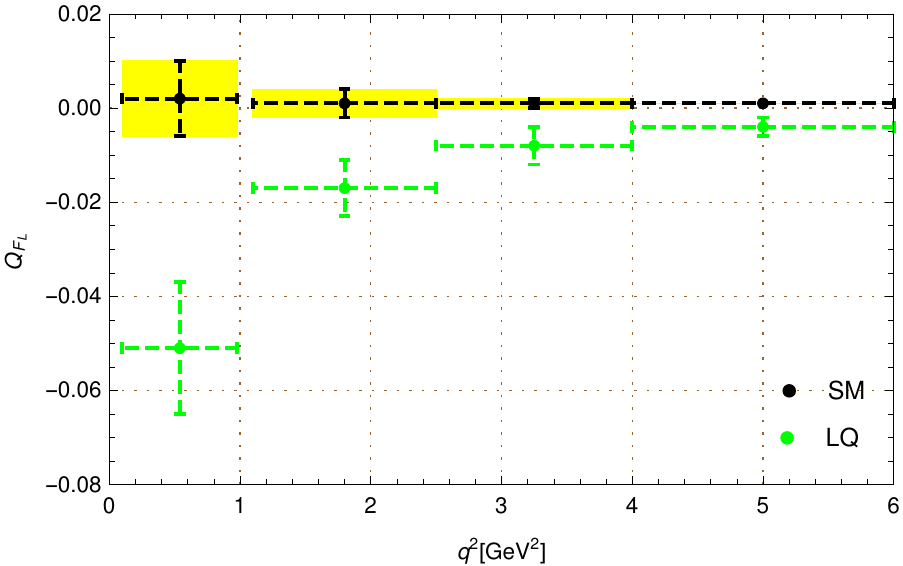}
\includegraphics[scale=0.63]{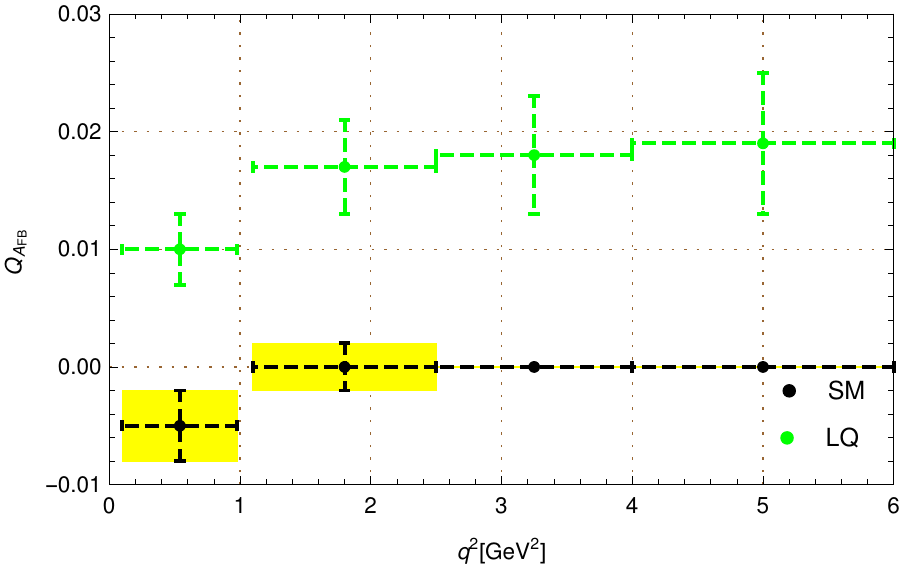}
\includegraphics[scale=0.63]{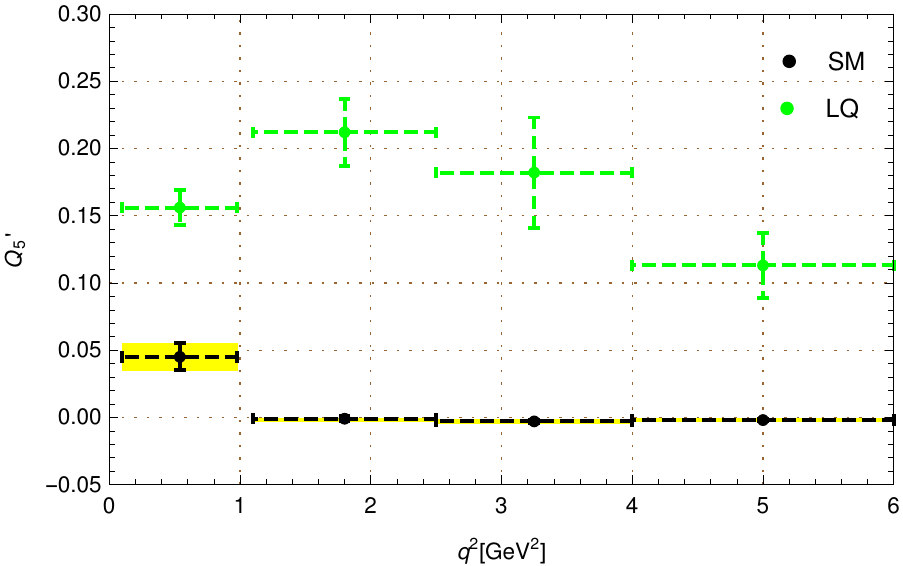}
\caption{\label{Bf2_Fig2}The bin wise distributions of observables such as branching ratio, the polarization fraction, the forward-backward asymmetry, $P_5^{\prime}$, and the sensitive LFU parameters $R_{f_2^{\prime}}$, $Q_{F_L}$, $Q_{A_{FB}}$ and $Q_5^{\prime}$ of $B_s \to f_2^{\prime} \ell ^+ \ell ^-$ processes (black bins: SM central values, yellow band: $1 \sigma$ uncertainty due to form factors and CKM element, green bin: $Z^{\prime}$ contribution).}
\end{figure}

We analyze the rare exclusive $B_{(s)} \to T \ell^+ \ell^-$ ($T= f_2^{\prime}, K_2^*$) processes in the presence of light $Z^{\prime}$ model where the coupling arises from only $C_9^{NP}$ contribution, in other words, the coupling correspond to the vectorial contribution to muon. Using the NP coupling, we report the impact on various observables such as differential branching ratio, the lepton polarization fraction $F_L$, the forward-backward asymmetry $A_{FB}$ and the angular observable $P_5^{\prime}$. Additionally some other important LFU sensitive observables such as $R_T$ ($T=f_2^{\prime}, K_2^*$), $Q_{F_L}$, $Q_{A_{FB}}$ and $Q_5^ {\prime}$ are also investigated in this analysis. With all the input parameters that are pertinent  to our analysis, we display the variations of all the observables w.r.t $q^2$ in Fig.~\ref{Bf2_Fig1}. Similarly in Fig.~\ref{Bf2_Fig2}, we show the corresponding $q^2$ bin wise plots for $B_s\to f_2^{\prime} \ell \ell$ decay mode where we choose different bin sizes such as [0.1,0.98], [1.1,2.5], [2.5,4], [4.0,6.0] and [1.1,6] (in the units of $\rm GeV^2$) compatible with LHCb measurements. The bin wise predictions along with its $1 \sigma$ standard deviation both in SM and in the presence of $Z^{\prime}$ model in several $q^2$ bin rooms have been reported in Table~\ref{tab_sm1}. We provide our detailed observations in the presence of NP contribution as below.

Description of the color inputs for the following plots:

Distribution plot :- black dotted line: SM contribution, cyan band: $1\sigma$ error band due to form factors and CKM element, orange dotted line: light $Z^{\prime}$ contribution.

Bin wise plot :- black bins: SM central values, yellow band: $1 \sigma$ uncertainty due to form factors and CKM element, green bin: light $Z^{\prime}$ contribution.

\begin{itemize}
\item Branching ratio (BR): 
In top-left panel of Fig.~\ref{Bf2_Fig1}, we show the $q^2$ dependency of the branching ratio for $B_s \to f_2 ^{\prime} \ell^+ \ell^-$ decay within SM as well as in the presence of light $Z'$ model for the $\mu$ mode. We observe that the $q^2$ behavior of the observable in the presence of light $Z'$ is reduced and lie within the SM $1 \sigma$ uncertainty band. Similarly, we proceed with the bin wise plot of the branching ratio in top-left panel of Fig.~\ref{Bf2_Fig2}. However, though the numerical values in the presence of light $Z^{\prime}$ differ from the SM contribution, no
such remarkable deviations are observed in this analysis.
\item Forward-backward asymmetry ($A_{FB}$): We display the $q^2$ variation of forward-backward asymmetry in the middle-left panel of Fig.~\ref{Bf2_Fig1}. In the presence of light $Z'$ contribution, it's $q^2$ behavior shifted to higher values as compared to SM variations in all bin rooms. In the SM variation, the observable $A_{FB}$ $(q^2)$ has  zero crossing at $\sim 2.8$ $\rm GeV^2$ whereas the crossing point shifted to $\sim 3.5$ $\rm GeV^2$ in the presence of new physics. Again we observe that in all bins given in the Table~\ref{tab_sm1}, the NP contributions lies within $1 \sigma$ from the SM predictions.
\item Longitudinal polarization fraction ($F_L$):  From the $q^2$ distribution plot given in middle-right panel of Fig.~\ref{Bf2_Fig1}, one can observe that due the NP coupling the contribution shifted lower to the SM values in all $q^2$  bins. However we do not draw any significant deviations for this observable.
\item The angular observable ($P_5 ^{\prime}$): For the angular observable $P_5^{\prime}$ given in the bottom panel of Fig.~\ref{Bf2_Fig1}, in the presence of NP coupling this observable is clearly distinguished from the SM contributions. However we observe that the NP coupling shift the contribution to higher values as compared to the SM. The zero crossing occurs at nearly $\sim 1.8$ $\rm GeV^2$ for the SM whereas in presence of NP coupling it touches at $\sim 2.3$ $\rm GeV^2$ for the same. This observable becomes negative in the $q^2$ regions [2.5,4], [4,6], and [1.1,6] whereas it remains positive in other bin ranges.
\item LFU sensitive parameter ($R_{f_2^{\prime}}$): Interestingly, the ratio of the branching ratio in other words the LFU sensitive parameter $R_{f_2^{\prime}}$ is clearly distinguishable from the SM prediction ($\simeq 1$) with more than $5 \sigma$ standard deviation in all bin ranges except $q^2 \in [0.1,0.98]$. The error band associated with this LFU parameter $R_{f_2^{\prime}}$ is almost zero.
\item The Q parameters ($\langle Q_{F_L} \rangle$, $\langle Q_{A_{FB}}\rangle$ and $\langle Q_5^{\prime} \rangle$): We provide the SM values and the NP contributions for each $q^2$ bin region in the bottom panel of Fig.~\ref{Bf2_Fig2} correspondingly. We observe that in all $Q_i$ ($Q_{F_L}$, $Q_{A_{FB}}$, $Q_5^{\prime}$) parameters, the predictions in the presence of light $Z^{\prime}$ deviates significantly from the SM values. For $Q_{F_L}$, specifically in the bin region [0.1,0.98] and [1.1,6.0], we get more than $3 \sigma$ standard deviation whereas in the rest of the bin rooms it is less than $3 \sigma$ from SM contribution. 
Similarly, in the $Q_{A_{FB}}$ observable, we get $(3 -5)$ $\sigma$ deviation in all $q^2$ bins. From the Table~\ref{tab_sm1}, one can observe clearly for another LFU parameter $Q_5^{\prime}$ that it varies $(4-9)$ $\sigma$ deviation from the SM in all bins enveloped in $q^2 \in [0.1,6.0]$.
\end{itemize}
\begin{table}[htbp]
\centering
\scalebox{1.1}{
\begin{tabular}{|c|c|c|c|c|c|c|}
\hline
Observable & & [0.10, 0.98] & [1.1, 2.5] & [2.5, 4.0] & [4.0, 6.0]&[1.1, 6.0] \\
\hline
\hline
\multicolumn{7}{|c|}{$B_s \to f_2^{\prime} \mu^+ \mu^-$}\\
\hline
\hline
\multirow{2}{*}{$\mathcal{BR}\times 10^{-7}$} 
& $SM$ & $0.344 \pm 0.107$ & $0.415 \pm 0.155$ & $0.408\pm 0.141$ & $0.464 \pm 0.153$ & $1.287\pm 0.449$  \\
\cline{2-7}
& $Z'$ & $0.296\pm 0.085$ & $0.345\pm 0.125$ & $0.337 \pm 0.114$ & $0.381 \pm 0.124$ & $1.064\pm 0.364$  \\
\hline $\mathcal{A}_\mathcal{FB}$ 
& $SM$ & $0.044 \pm 0.012$ & $0.025\pm 0.021$ & $-0.014\pm 0.012$ & $-0.061 \pm 0.016$ & $-0.018\pm 0.014$  \\
\cline{2-7}
& $Z'$ & $0.055\pm 0.013$ & $0.043 \pm 0.021$ & $0.004 \pm 0.014$ & $-0.041 \pm 0.014$ & $0.000\pm 0.013$  \\
\hline
\multirow{2}{*}{$\mathcal{F}_\mathcal{L}$} 
& $SM$ & $0.782 \pm 0.087$ & $0.951\pm 0.028$ & $0.928\pm 0.012$ & $0.871 \pm 0.011$ & $0.915\pm 0.013$  \\
\cline{2-7}
& $Z'$ & $0.728\pm 0.103$ & $0.932 \pm 0.028$ & $0.918\pm 0.012$ & $0.865 \pm 0.017$ & $0.903\pm 0.013$  \\
\hline
\multirow{2}{*}{$\mathcal{P}^{\prime}_5$} 
& $SM$ & $0.649\pm 0.110$ & $0.011 \pm 0.188$ & $-0.451\pm 0.145$ & $-0.576 \pm 0.118$ & $-0.381\pm 0.140$  \\
\cline{2-7}
& $Z'$ & $0.759\pm 0.103$ & $0.224\pm 0.152$ & $-0.265 \pm 0.140$ & $-0.460 \pm 0.120$ & $-0.209\pm 0.149$  \\
\hline
\multirow{2}{*}{$\mathcal{R}^{\mu}_e$} 
& $SM$ & $0.984 \pm 0.039$ & $0.996\pm 0.018$ & $0.997\pm 0.005$ & $0.997 \pm 0.002$ & $0.997\pm 0.006$  \\
\cline{2-7}
& $Z'$ & $0.846\pm 0.060$ & $0.829 \pm 0.021$ & $0.822 \pm 0.009$ & $0.820 \pm 0.006$ & $0.823\pm 0.010$  \\
\hline
\multirow{2}{*}{$Q_{\mathcal{A}_\mathcal{FB}}$} 
& $SM$ & $-0.005 \pm 0.003$ & $-0.000 \pm 0.002$ & $0.000\pm 0.000$ & $0.000 \pm 0.000$ & $-0.000\pm 0.000$  \\
\cline{2-7}
& $Z'$ & $0.010\pm 0.003$ & $0.017\pm 0.004$ & $0.018 \pm 0.005$ & $0.019 \pm 0.006$ & $0.018\pm .0.004$  \\
\hline
\multirow{2}{*}{$Q_{\mathcal{F}_\mathcal{L}}$} 
& $SM$ & $0.002 \pm 0.008$ & $0.001\pm 0.003$ & $0.001\pm 0.001$ & $0.001 \pm 0.000$ & $0.001\pm 0.000$  \\
\cline{2-7}
& $Z'$ & $-0.051\pm 0.014$ & $-0.017\pm 0.006$ & $-0.008\pm 0.004$ & $-0.004 \pm 0.002$ & $-0.010\pm 0.003$  \\
\hline
\multirow{2}{*}{$Q^{\prime}_5$} 
& $SM$ & $0.045 \pm 0.010$ & $-0.001\pm 0.002$ & $-0.003\pm 0.001$ & $-0.002 \pm 0.001$ & $-0.004\pm 0.001$  \\
\cline{2-7}
& $Z'$ & $0.156\pm 0.013$ & $0.212 \pm 0.025$ & $0.182 \pm 0.041$ & $0.113 \pm 0.024$ & $0.166\pm 0.037$  \\
\hline
\hline
\end{tabular}
}
\caption{Prediction of various observables with $1 \sigma$ standard deviation in SM and $Z^{\prime}$ model for the $ B_s \to f_2^{\prime}\ell^+ \ell^-$ process in different bin rooms}
\label{tab_sm1}
\end{table}

\newpage
\subsubsection*{Analysis of $B \to K_2^*(1430) \ell^+ \ell^-$ in SM and beyond}
\begin{figure}[htb]
\centering
\includegraphics[scale=0.7]{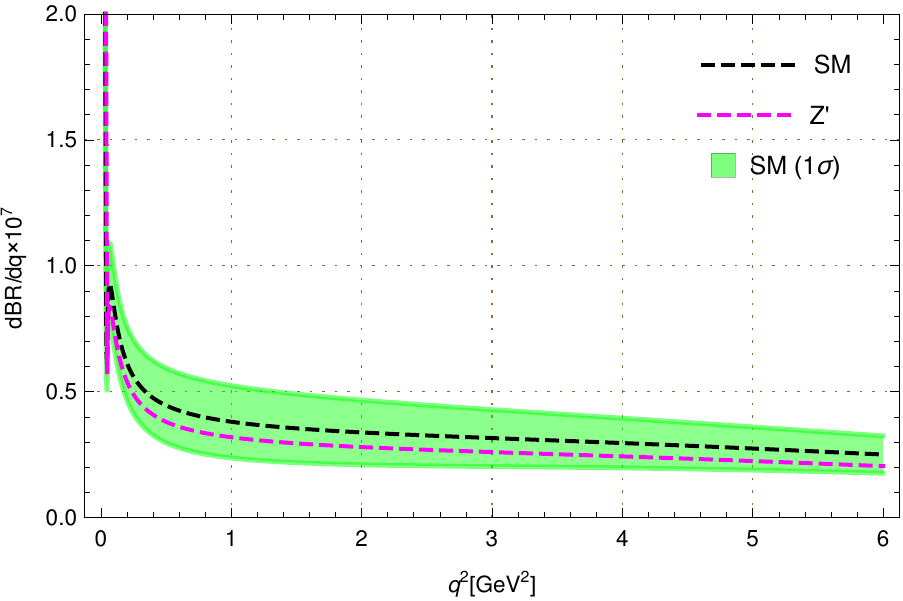} 
\quad
\includegraphics[scale=0.7]{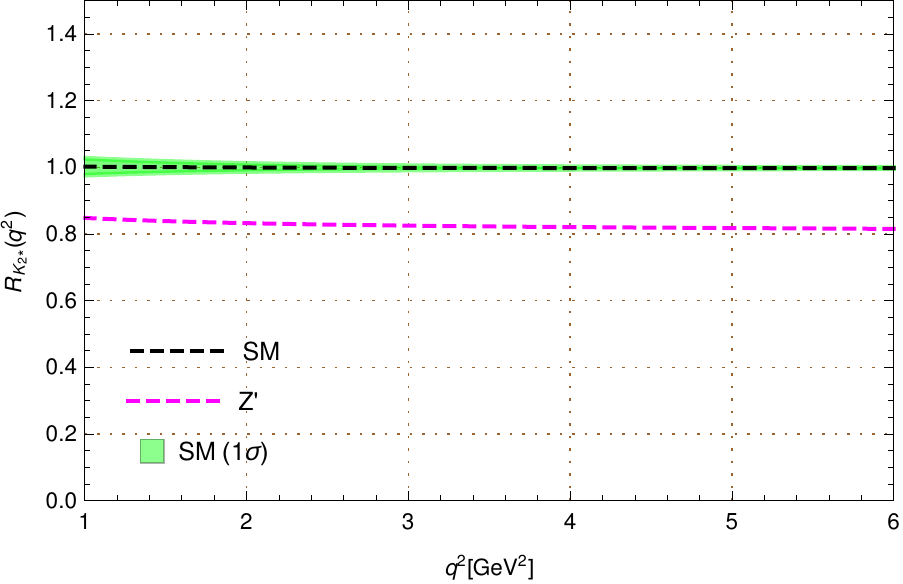} 
\quad
\includegraphics[scale=0.7]{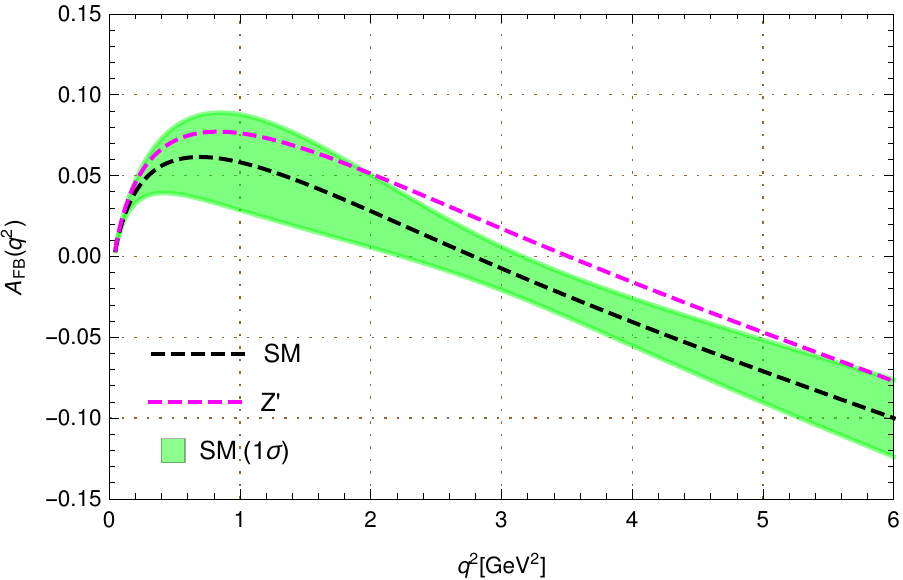} 
\quad
\includegraphics[scale=0.7]{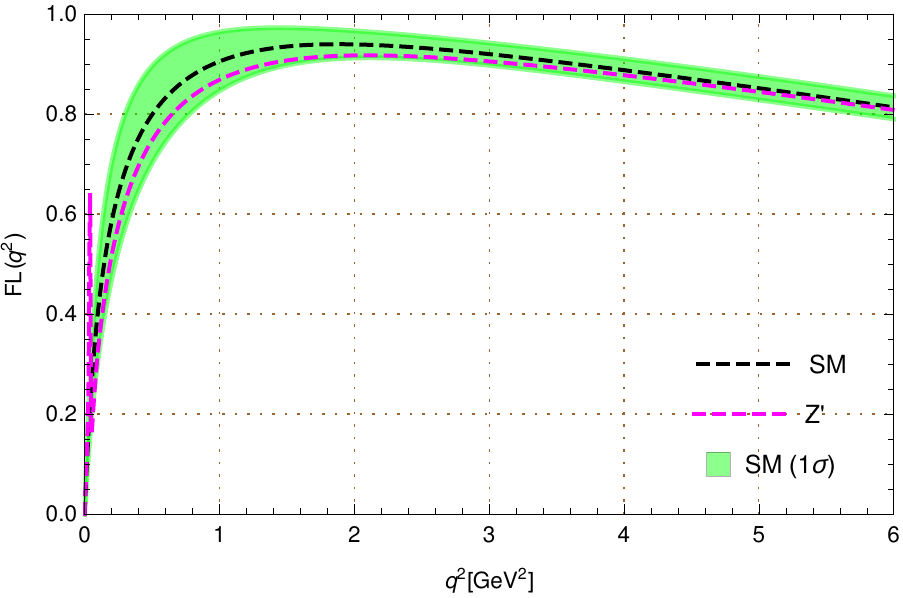}
\quad
\includegraphics[scale=0.7]{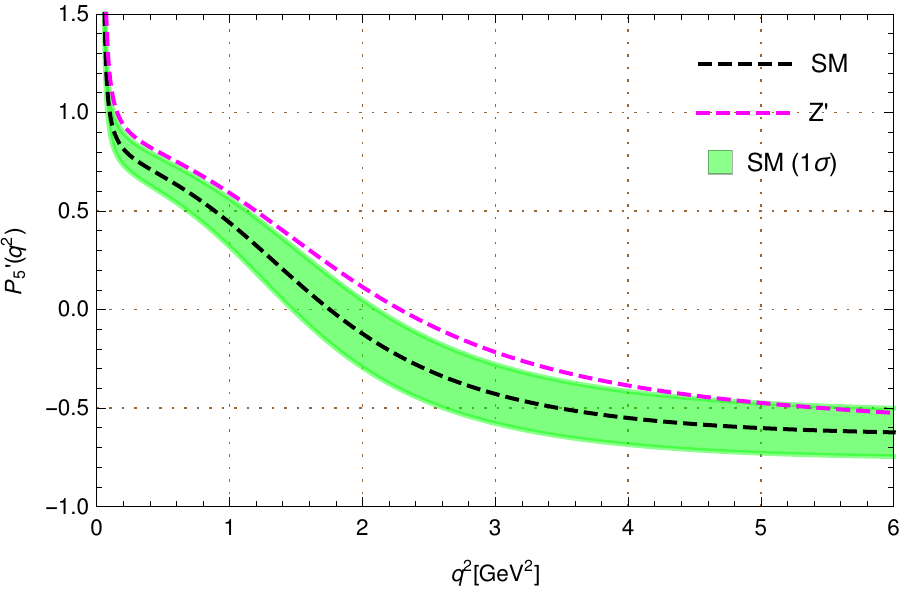}
\caption{The $q^2$ distribution of various observables such as branching ratio, the polarization fraction, the forward-backward asymmetry and $P_5^{\prime}$ for $B_s \to K_2^*(1430) \ell ^+ \ell ^-$ process (black dotted line: SM contribution, green band: $1 \sigma$ uncertainty due to form factors and CKM element, magenta dotted line: $Z^{\prime}$ contribution).}
\label{BK2_Fig1}
\end{figure}

\begin{figure}[htb]
\centering
\includegraphics[scale=0.63]{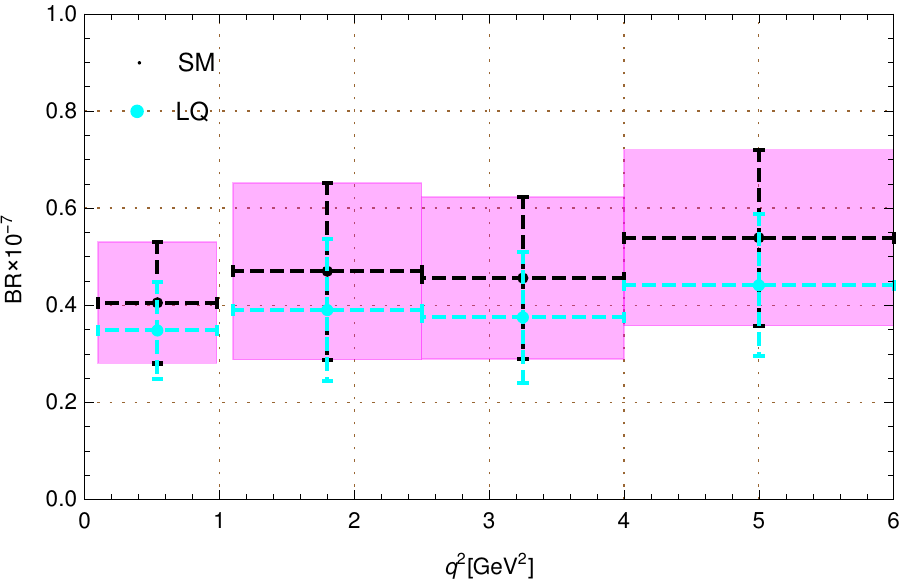} 
\includegraphics[scale=0.63]{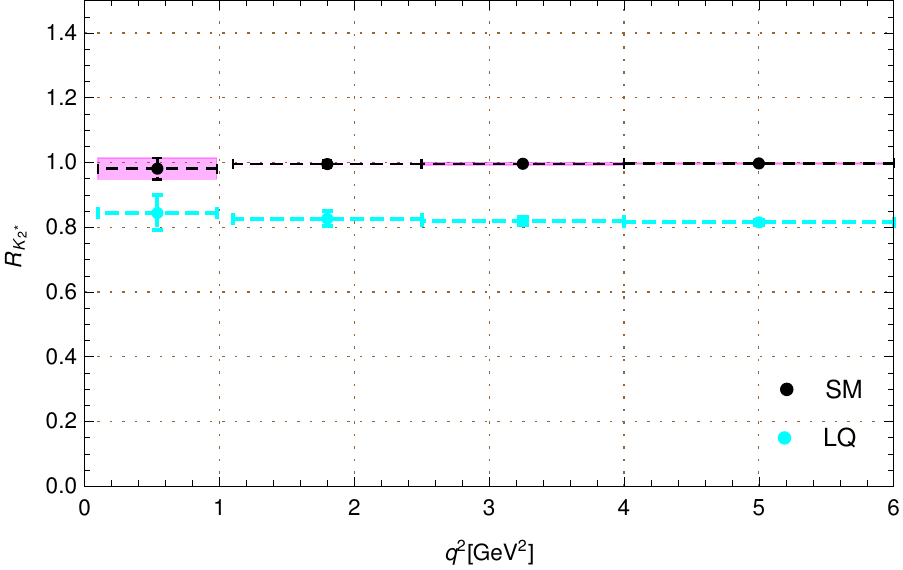} 
\includegraphics[scale=0.63]{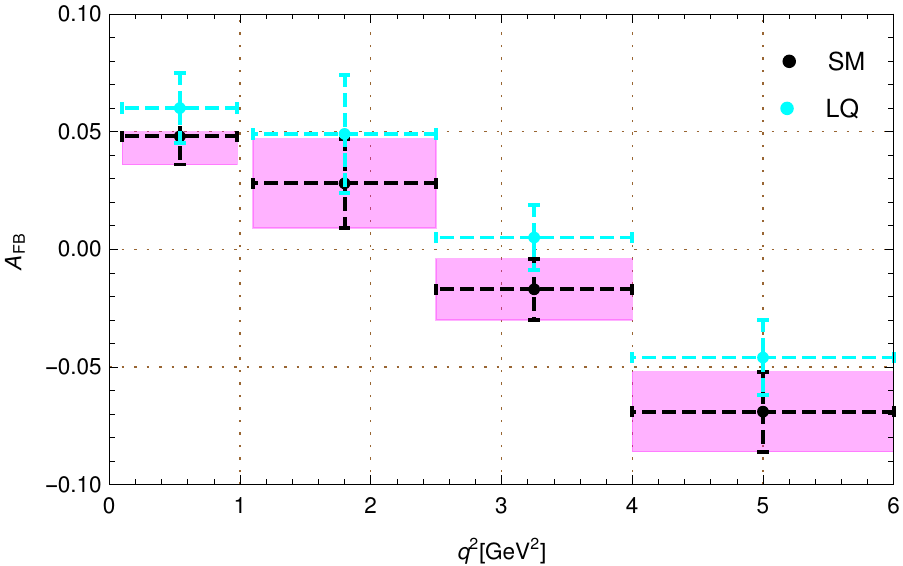} 
\includegraphics[scale=0.63]{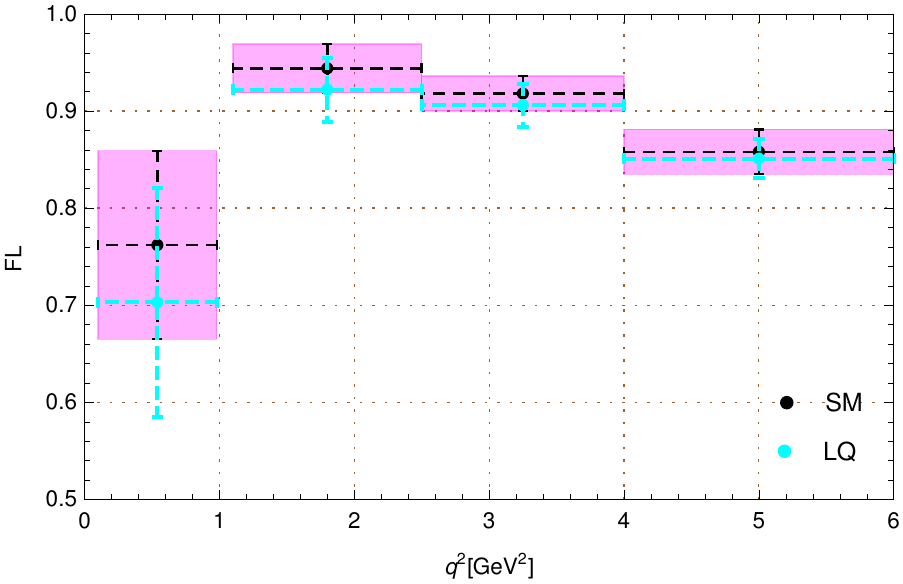}
\includegraphics[scale=0.63]{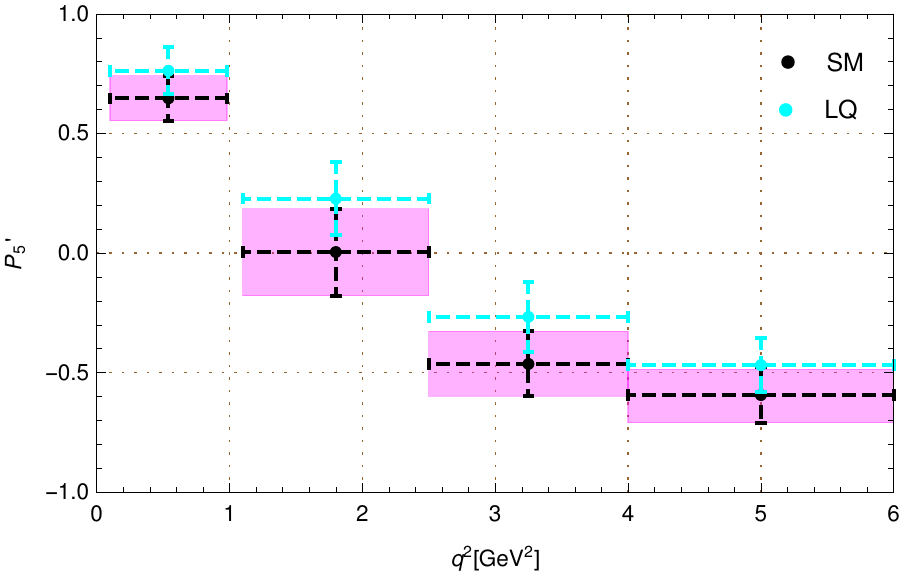}

\includegraphics[scale=0.63]{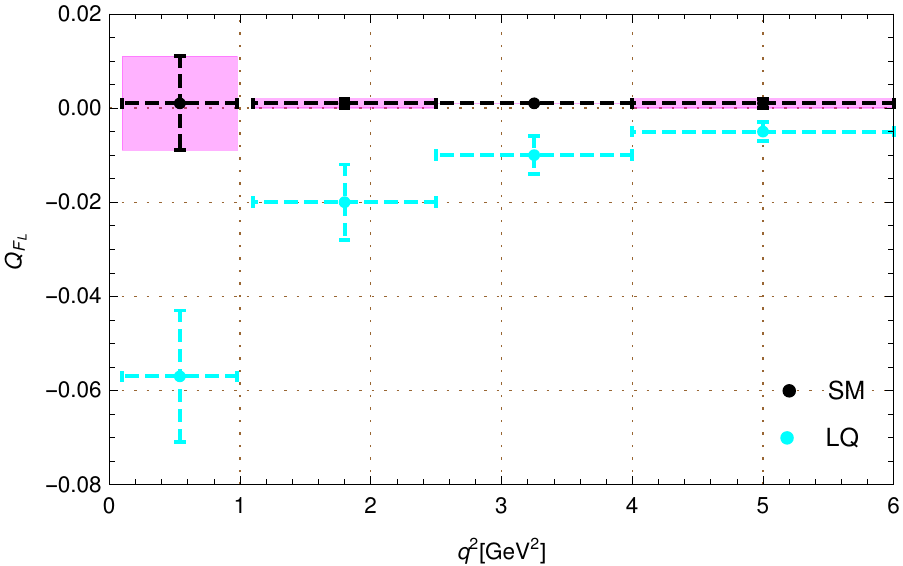} 
\includegraphics[scale=0.63]{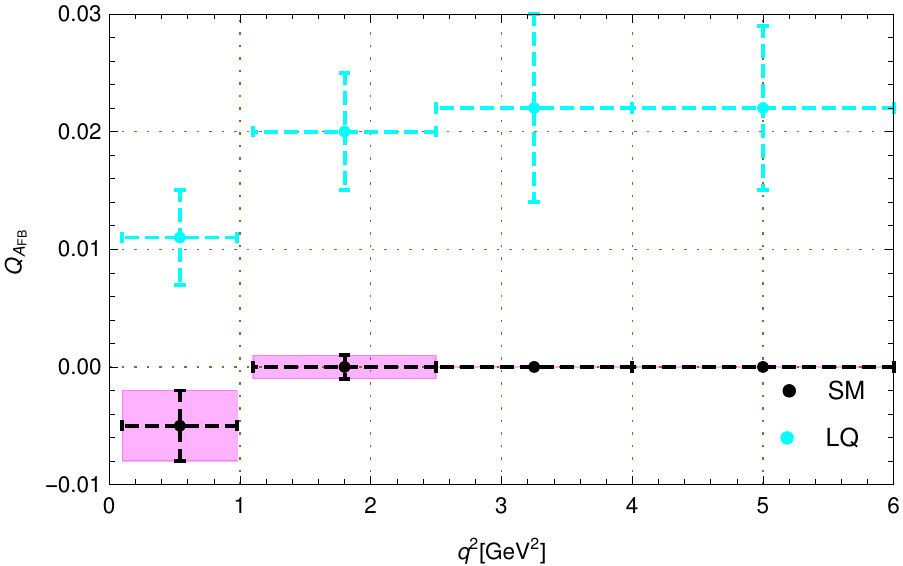}
\includegraphics[scale=0.63]{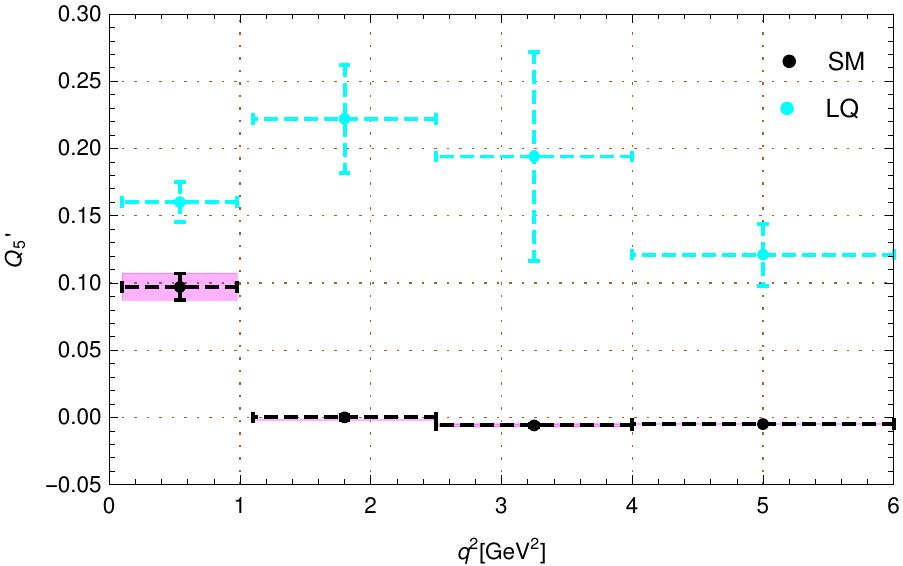}
\caption{\label{BK2_Fig2}The bin wise distributions of observables such as branching ratio, the polarization fraction, the forward-backward asymmetry, $P_5^{\prime}$, and the sensitive LFU parameters $R_{K_2^*}$, $Q_{F_L}$, $Q_{A_{FB}}$ and $Q_5^{\prime}$ of $B \to K_2^* \ell ^+ \ell ^-$ processes (black bins: SM central values, magenta band: $1 \sigma$ uncertainty due to form factors and CKM element, cyan bin: $Z^{\prime}$ contribution).}
\end{figure}

Similar to $B_s \to f_2^{\prime} \ell^+ \ell^-$ process, we also probe the semileptonic B meson decay to another tensor meson $K_2^*(1430)$ in the final state which also mediate $b \to s\ell \ell$ flavor changing neutral current transition. Here we also study the variation of the various observables such as BR, $F_L$, $A_{FB}$, $P_5^{\prime}$ and the LFU sensitive observables $R_{K_2^*}$, $Q_{F_L}$, $Q_{A_{FB}}$ and $Q_5^{\prime}$ both in SM as well as in the presence of light $Z^{\prime}$ model in Fig.~\ref{BK2_Fig1} where $1\sigma$ error to the SM contribution due to form factor and CKM element have been considered. In addition to this we display the corresponding bin plots in Fig.~\ref{BK2_Fig2} . We report the numerical results for all the observables at different $q^2$ bin regions in Table~\ref{tab_sm2}. We give details of our inspection as below.

Description of the color inputs for the following plots:

Distribution plot :- black dotted line: SM contribution, green band: $1\sigma$ error band due to form factors and CKM element, magenta dotted line: light $Z^{\prime}$ contribution.

Bin wise plot :- black bins: SM central values, magenta band: $1 \sigma$ uncertainty due to form factors and CKM element, cyan bin: light $Z^{\prime}$ contribution.
\begin{itemize}
\item Branching ratio (BR): We observe the $q^2$ behavior in the differential branching ratio of $B \to K_2^* \ell^+ \ell^-$ process both in SM as well as in the NP scenario that displayed in the top-left panel of Fig.~\ref{BK2_Fig1}. Not being significant,  the observable in the presence of the NP coupling is reduced in comparison to the SM values. In all bin regions, the observable spans less than $1\sigma$ deviation from the SM predictions.
\item Forward-backward asymmetry ($A_{FB}$): We observe the zero crossing point of the observable $A_{FB}(q^2)$ in the SM at  $\sim 2.8 \rm GeV^2$ whereas it shifted to higher value at $\sim 3.5 \rm GeV^2$ in the presence of NP coupling. The light $Z^{\prime}$ contribution is clearly distinguishable in the range $q^2 \in$ [2.5,4]  and [1.1,6] with $1.15 \sigma$ and $1.09 \sigma$ significance  respectively whereas less than $1 \sigma$ deviation is observed in the rest of the bin regions. 
\item Longitudinal polarization fraction ($F_L$): In the middle-right panel of Fig.~\ref{BK2_Fig1}, the $q^2$ dependency of the longitudinal polarization fraction $F_L(q^2)$ suddenly  increases up to the peak value at $\sim 1.4 \rm GeV^2$ and then decreases accordingly as $q^2$ value increases. However, it is observed that the peak of the observable in light $Z^{\prime}$ reduces and shifted to lower value than the SM contribution. Here also no remarkable deviation has been observed in the presence of NP scenario.
\item $P_5 ^{\prime}$: The angular observable $P_5^{\prime}$ is also $q^2$ dependent and is clearly provide a remarkable contribution in the presence of NP coupling. It is observed that the zero crossing point in the SM is at $\sim 1.75$ $ \rm GeV^2$ whereas the $Z^{\prime}$ contribution shift this point to higher value at $\sim 2.30$ $ \rm GeV^2$. 
\item LFU sensitive parameter ($R_{K_2^*}$, $\langle Q_{F_L} \rangle$, $\langle Q_{A_{FB}}\rangle$ and $\langle Q_5^{\prime} \rangle$): In the case of the LFU sensitive parameter $R_{K_2^*}$ shown in the top-right panel of Fig.~\ref{BK2_Fig1}, the observable is quite distinguishable in the presence of light $ Z^{\prime}$ scenario. However, we observe more than $5 \sigma$ deviation than SM contribution in all $q^2$ bin regions starting from 1.1 to 6 $\rm GeV^2$ whereas in the range $q^2 \in [0.1,0.98]$, $1.92 \sigma$ significance is observed for this observable.\\
Like $R_{K_2^*}$, the $Q_i$ parameters significantly deviate from the SM. For $Q_{A_{FB}}$, the $Z^{\prime}$ contribution provide $(2-4)$ $\sigma$ deviation in all bin regions as compared to SM contribution. Similarly, we notice $3.37 \sigma$ standard deviation in the bin range [0.1,0.98] and $< 3 \sigma$ in all other $q^2$ bin ranges for the observable $Q_{F_L}$. Last but not the least, the parameter $Q_5^{\prime}$ can be observed with more than $5 \sigma$ in the region $q^2 \in \big([1.1,2.5]$ , $[4,6] \big)$ whereas $3.49 \sigma$ and $3.518 \sigma$ in the regions [0.1,0.98] and [1.1,6.0] respectively. However, in the $q^2 \in [2.5,4]$ region, we get 2.56 $\sigma$ deviation from the SM contribution. The bin wise plots for all the above discussed observables are shown in Fig.~\ref{BK2_Fig2}.
\end{itemize}

\begin{table}[htbp]
\centering
\label{tab_sm2}
\scalebox{1.1}{
\begin{tabular}{|c|c|c|c|c|c|c|}
\hline
Observable & & [0.10, 0.98] & [1.1, 2.5] & [2.5, 4.0] & [4.0, 6.0]&[1.1, 6.0] \\
\hline
\hline
\multicolumn{7}{|c|}{$B \to K_2^* \mu^+ \mu^-$}\\
\hline
\hline
\multirow{2}{*}{$\mathcal{BR}\times 10^{-7}$} 
& $SM$ & $0.405 \pm 0.125$ & $0.470 \pm 0.182$ & $0.456\pm 0.167$ & $0.539 \pm 0.181$ & $1.467 \pm 0.531$  \\
\cline{2-7}
& $Z'$ & $0.348\pm 0.099$ & $0.390\pm 0.147$ & $0.375 \pm 0.135$ & $0.441 \pm 0.146$ & $1.208 \pm 0.428$  \\
\hline $\mathcal{A}_\mathcal{FB}$ 
& $SM$ & $0.048 \pm 0.012$ & $0.028 \pm 0.019$ & $-0.017\pm 0.013$ & $-0.069 \pm 0.017$ & $-0.021 \pm 0.012$  \\
\cline{2-7}
& $Z'$ & $0.060\pm 0.015$ & $0.049\pm 0.025$ & $0.005\pm 0.014$ & $-0.046 \pm 0.016$ & $0.000 \pm 0.015$  \\
\hline
\multirow{2}{*}{$\mathcal{F}_\mathcal{L}$} 
& $SM$ & $0.762 \pm 0.097$ & $0.944 \pm 0.025$ & $0.918\pm 0.018$ & $0.858 \pm 0.023$ & $0.904 \pm 0.026$  \\
\cline{2-7}
& $Z'$ & $0.703\pm 0.118$ & $0.922 \pm 0.033$ & $0.906 \pm 0.022$ & $0.851 \pm 0.020$ & $0.891 \pm 0.030$  \\
\hline
\multirow{2}{*}{$\mathcal{P}^{\prime}_5$} 
& $SM$ & $0.647 \pm 0.094$ & $0.004 \pm 0.182$ & $-0.464\pm 0.136$ & $-0.594 \pm 0.116$ & $-0.395 \pm 0.128$  \\
\cline{2-7}
& $Z'$ & $0.762 \pm 0.098$ & $0.227\pm 0.153$ & $-0.267 \pm 0.147$ & $-0.469 \pm 0.112$ & $-0.213\pm 0.145$  \\
\hline
\multirow{2}{*}{$\mathcal{R}^{\mu}_e$} 
& $SM$ & $0.981\pm 0.033$ & $0.995 \pm 0.011$ & $0.996\pm 0.004$ & $0.997 \pm 0.002$ & $0.996 \pm 0.006$  \\
\cline{2-7}
& $Z'$ & $0.844 \pm 0.054$ & $0.826\pm 0.023$ & $0.819 \pm 0.011$ & $0.815 \pm 0.007$ & $0.820 \pm 0.008$  \\
\hline
\multirow{2}{*}{$Q_{\mathcal{A}_\mathcal{FB}}$} 
& $SM$ & $-0.005 \pm 0.003$ & $-0.000 \pm 0.001$ & $0.000\pm 0.000$ & $0.000 \pm 0.000$ & $-0.000\pm 0.000$  \\
\cline{2-7}
& $Z'$ & $0.011 \pm 0.004$ & $0.020 \pm 0.005$ & $0.022 \pm 0.008$ & $0.022 \pm 0.007$ & $0.022 \pm 0.008$  \\
\hline
\multirow{2}{*}{$Q_{\mathcal{F}_\mathcal{L}}$} 
& $SM$ & $0.001 \pm 0.010$ & $0.001 \pm 0.001$ & $0.001 \pm 0.000$ & $0.001 \pm 0.001$ & $0.001 \pm 0.001$  \\
\cline{2-7}
& $Z^{\prime}$ & $ -0.057\pm 0.014$ & $-0.020\pm 0.008$ & $ -0.010\pm 0.004$ & $-0.005 \pm 0.002$ & $-0.011 \pm 0.005$  \\
\hline
\multirow{2}{*}{$Q^{\prime}_5$} 
& $SM$ & $0.097 \pm 0.010$ & $-0.000 \pm 0.002$ & $-0.006 \pm 0.001$ & $-0.005 \pm 0.000$ & $-0.006\pm 0.001$  \\
\cline{2-7}
& $Z^{\prime}$ & $ 0.160\pm 0.015$ & $0.222 \pm 0.040$ & $0.194 \pm 0.078 $ & $0.121\pm 0.023$ & $0.177 \pm 0.052$  \\
\hline
\hline
\end{tabular}
}
\caption{Prediction of various observables with $1 \sigma$ standard deviation in SM and $Z^{\prime}$ model for the $ B_s \to f_2^{\prime}\ell^+ \ell^-$ process in different bin rooms}
\label{tab_sm2}
\end{table}


\section{Conclusion}\label{conc}
Inspired by the anomalies present in $B \to (K, K^*) \ell^+ \ell^-$ and $B_s \to \phi \mu^+ \mu^-$ decays proceeding via $b\to s \ell^+ \ell^-$ flavor changing neutral current quark level interaction,  we scrutinize the semileptonic decays of $B \to K_2^* (1430)$ and $B_s \to f_2 ^{\prime} (1525)$ with the charged leptons ($\ell = \mu, e)$ in the presence of SM and the light $Z^{\prime}$ model. Assuming the NP present in muon mode of lepton pair in the final state, we constrain the NP coupling by considering the experimental data associated with the clean observable $R_K$ in the range $1.1 <q^2<6.0$ $\rm GeV^2$ and $R_{K^*}$ in the central $q^2$ region [1.1,6.0] with the performance of $\chi^2$ fit. In the presence of effective Hamiltonian for $b \to s \ell \ell$ transition, we provide a detailed study of the behavior of various physical observables such as differential branching ratio, lepton polarization fraction, forward-backward asymmetry, the angular observable $P_5^{\prime}$, and LFU sensitive parameter as the ratio of branching ratios in $B\to K_2^*$ and $B_s \to f_2^{\prime}$ transition with $\mu$ mode to e mode in the final state in the SM as well as in the presence of light $Z^{\prime}$. The other observables that are very sensitive to lepton flavor universality also draw attention to probe on few $Q_i$ parameters corresponding to the longitudinal polarization fraction ($Q_{F_L}$), forward-backward asymmetry ($Q_{A_{FB}}$) and the angular observable $P_5 ^{\prime}$ ($Q_5^ {\prime}$). With the $q^2$ dependent NP coupling we give the integrated predictions of all the above discussed prominent observables pertaining to $B\to K_2^* \ell^+ \ell^-$ and $B_s \to f_2^{\prime} \ell^+ \ell^-$ decays at different $q^2$ bin regions that compatible with LHCb experiment. In this study all the observables  are investigated by considering the form factors obtained from light cone sum rule approach. 

We observed in our analysis that the differential branching ratio is reduced as compared to SM and notice that no significant deviation for this observable in both exclusive $B\to K_2^* \ell^+ \ell^-$ and $B_s \to f_2^{\prime} \ell^+ \ell^-$ processes in the presence of light $Z^{\prime}$ boson. In the observables the longitudinal fraction and the angular observable $P_5^{\prime}$, we get a remarkable contribution in the new physics analysis in both the decay modes. The deviations observed at the LFU parameters such as $R_{f_2^{\prime}}$ and $R_{K_2^*}$ are clearly distinguishable and as a complementary decay channel, both can provide an insight into the $R_{f_2^{\prime}}$ and $R_{K_2^*}$ anomalies which could be observed in the LHCb experiment. On the other hand, we also look into the $Q_i$ parameters which are very sensitive to LFUV and found that all the observables have profound deviations from the SM contribution. As the $B\to K_2^* \ell^+ \ell^-$ and $B_s \to f_2^{\prime} \ell^+ \ell^-$ decay processes have received less attention unlike $B \to (K, K^*) \ell^+ \ell^-$ and $B_s \to \phi \mu^+ \mu^-$ decays mediated by $b\to s \ell \ell$ quark level transition , it is very important to acquire more data sample from the experiments in order to understand the significance of new physics contributions.


\acknowledgments 
MKM would like to acknowledge DST INSPIRE fellowship division, Government of India for the financial support with ID - IF160303.

\FloatBarrier

\appendix


\end{document}